\documentclass[times,final,authoryear]{elsarticle} 
\usepackage[english]{babel}

\usepackage[letterpaper,top=2cm,bottom=2cm,left=3cm,right=3cm,marginparwidth=1.75cm]{geometry}

\usepackage{bm}
\usepackage{amsmath}
\usepackage{amssymb}
\usepackage{setspace}
\usepackage{mathalfa}
\usepackage{graphicx}
\usepackage{float}
\usepackage[colorlinks=true, allcolors=blue]{hyperref}
\usepackage{cleveref}
\usepackage[table,xcdraw]{xcolor}
\usepackage{caption}
\usepackage{subcaption}
\usepackage{tikz}
\usetikzlibrary{decorations, positioning, arrows.meta, shapes.geometric}
\usepackage{pgfplots}
\pgfplotsset{compat=newest}
\usepackage{algorithm}
\usepackage{algorithmic}
\usepackage{xcolor}
\usepackage{transparent}
\definecolor{darkred}{RGB}{200, 0, 0}

\newcommand{\SPL}{\text{SPL}} % Shortcut 
\newcommand{\transparency}[2][0.5]{\transparent{#1}#2\transparent{1}}
\definecolor{python_blue}{rgb}{0,0.10980392156862745,0.4980392156862745}
\definecolor{python_green}{rgb}{0.00392156862745098,0.4588235294117647,0.09019607843137255}
\definecolor{python_red}{rgb}{0.5490196078431373,0.03529411764705882,0}
\definecolor{python_purple}{rgb}{0.4627450980392157,0,0.6313725490196078}
\begin{document}

\begin{frontmatter}
\title{Deep Reinforcement Learning for Multi-Objective Optimization: Enhancing Wind Turbine Energy Generation while Mitigating Noise Emissions 
}
\author[1]{Mart\'in de Frutos\corref{cor1}}
\cortext[cor1]{Corresponding author}
\ead{m.defrutos@upm.es}
\author[1]{Oscar A. Marino}
\author[1]{David Huergo}
\author[1,2]{Esteban Ferrer}

\address[1]{ETSIAE-UPM-School of Aeronautics, Universidad Politécnica de Madrid, Plaza Cardenal Cisneros 3, E-28040 Madrid, Spain}
\address[2]{Center for Computational Simulation, Universidad Politécnica de Madrid, Campus de Montegancedo, Boadilla del Monte, 28660 Madrid, Spain}

\begin{keyword}
Wind turbine \sep Deep Reinforcement Learning\sep  Q-learning \sep Blade Element Momentum Theory \sep Aeroacoustic \sep Brooks Pope and Marcolini  \sep multi-objective optimization \sep torque-pitch control
\end{keyword}

\begin{abstract}
We develop a torque-pitch control framework using deep reinforcement learning for wind turbines to optimize the generation of wind turbine energy while minimizing operational noise. We employ a double deep Q-learning, coupled to a blade element momentum solver, to enable precise control over wind turbine parameters. In addition to the blade element momentum, we use the wind turbine acoustic model of Brooks Pope and Marcolini. Through training with simple winds, the agent learns optimal control policies that allow efficient control for complex turbulent winds. Our experiments demonstrate that the reinforcement learning is able to find optima at the Pareto front, when maximizing energy while minimizing noise. In addition, the adaptability of the reinforcement learning agent to changing turbulent wind conditions, underscores its efficacy for real-world applications. We validate the methodology using a SWT2.3-93 wind turbine with a rated power of 2.3 MW. We compare the reinforcement learning control to classic controls to show that they are comparable when not taking into account noise emissions. When including a maximum limit of 45 dB to the noise produced (100 meters downwind of the turbine), the extracted yearly energy decreases by 22\%. The methodology is flexible and allows for easy tuning of the objectives and constraints through the reward definitions, resulting in a flexible multi-objective optimization framework for wind turbine control.
Overall, our findings highlight the potential of RL-based control strategies to improve wind turbine efficiency while mitigating noise pollution, thus advancing sustainable energy generation technologies.
\end{abstract}

\end{frontmatter}

% \tableofcontents

\section{Introduction}
\label{sec:introduction}
In recent years, the aerodynamic design of wind turbines has undergone significant advances, reaching near-optimal efficiency through substantial investments in aerodynamic optimization, as well as advancements in manufacturing techniques and materials. Consequently, the focus has shifted towards addressing the issue of noise generated by wind turbines, which is emerging as a competitive factor within the wind energy industry. Numerous studies have explored the correlation between wind turbine sound power levels and public-reported perceptions of annoyance \citep{wolsink1993annoyance,pedersen2009response}. Accurate prediction of wind turbine noise under real operational and atmospheric conditions is crucial to design quieter turbines and complying with imposed noise regulations \citep{DAVY2018288}. This necessity underscores the importance of fast turn-around methods to incorporate noise calculations into design and optimization processes, as well as to assess noise in real time during operation. Such efforts not only optimize wind resource utilization, but also minimize the impact on the quality of life of nearby communities and wildlife.

Aerodynamic noise poses a significant limitation to further exploiting wind energy resources. This type of noise results from the turbulent flow interacting with the airframe, necessitating a detailed resolution of the flow for accurate far-field noise prediction. However, computational fluid dynamics (CFD) solvers, while capable of simulating the flow field, incur a high computational cost which escalates further when resolving the acoustic field. Consequently, numerical approaches to wind turbine noise prediction remain challenging. Therefore, most noise prediction models for wind turbines are based on aeroacoustic semi-empirical models rather than numerical simulations \citep{wagner1996noise}. Despite these obstacles, wind turbines remain an essential component in the generation of clean and renewable energy. However, effective control strategies are imperative to optimize their performance under variable wind conditions.

Wind turbine control systems are designed to maximize energy generation while ensuring structural integrity and safe operation \citep{NJIRI2016377,NOVAESMENEZES2018945,https://doi.org/10.1049/rpg2.12160}. Given the dynamic nature of wind, adaptive control strategies are essential, with classic mechanisms including adjustments to yaw angle and rotational speed, as well as blade pitch angle modulation. Leveraging real-time wind measurements, turbine dynamics, and advanced control algorithms enables simultaneous adjustments to rotor speed, and pitch, enhancing energy generation, reducing fatigue loads, and extending turbine lifespan.

The emergence of reinforcement learning (RL) presents novel opportunities for wind turbine control by enabling data-driven adaptive decision-making \citep{LECLAINCHE2023108354,GARNIER2021104973}. RL, a machine learning approach, involves an agent learning to make decisions in an environment to maximize cumulative rewards over time \citep{sutton1998rli}. Applied to wind turbines, RL offers autonomous learning of control inputs to maximize power generation by capturing complex non-linear relationships between wind conditions, turbine states, and actions. RL-based control methods adapt in real-time to changing wind conditions, offering significant advantages in wind turbine operation.

This paper reviews recent advances in RL-based control strategies for wind turbines, focusing on pitch angle and rotor speed modulation. Previous studies have proposed RL algorithms with comprehensive reward definitions, showcasing their efficacy in optimizing wind turbine performance under varying wind conditions. %\citep{SESMPR20,SESM20,XIE2023118893,chen2020reinforcement,KADOCHE2023119129,PARJ23,kushwaha2020q,wei2016adaptive,SAZEFULJLJ19,2SAZEFULJLJ19,DHZJZX21}.
For example \cite{chen2020reinforcement} proposed a RL pitch controller to maintain the nominal power using an adaptive dynamic programming algorithm, reducing the energy consumption of the pitch actuator. \cite{XIE2023118893} developed a data-driven model to perform a torque-pitch controller, modeling the dynamics and using RL to control the wind turbine. \cite{SESMPR20} discussed different reward definitions for wind turbine controllers, while \cite{PARJ23} and \cite{SAZEFULJLJ19} developed RL methods for yaw control avoiding control parameter tuning. \cite{kushwaha2020q} and \cite{wei2016adaptive} employed Q-learning RL methods for maximum power point tracking (MPPT) control of the generator speed. Overall, these studies demonstrate the adaptability of RL systems to realistic wind conditions, thereby enhancing overall energy generation and efficiency of wind farms.

In this paper, we introduce a reinforcement learning-based dynamic control method designed to maximize power output while adhering to specified maximum decibel levels. The paper is structured as follows. First, we summarize the methodology in \Cref{sec:methodology}. There, we include the wind turbine model, validating the aeroacoustic model with three different wind conditions. Additionally, the multi objective reinforcement learning strategy is explained, we provide details on the reward, the neural network architecture and the training procedure. Second, in \Cref{sec:results} we validate the controller with simple steady winds to later challenge the method with turbulent wind conditions obtained from experimental measurements. We end with conclusions and outlooks.
\section{Methodology}
\label{sec:methodology}
In this section, we detail the methodology for integrating Deep Reinforcement Learning (DRL) with the dynamic control of a wind turbine. We begin by describing the model of the wind turbine, focusing on how both the power output and the noise levels are computed, and validate the methodology using field measurements for a SWT2.3-93 wind turbine with a rated power of 2.3 MW. Subsequently, we detail the setup of the DRL algorithm, which is designed to maximize power generation within specified noise constraints, demonstrating the application of advanced machine learning techniques to real-world energy optimization challenges.
\subsection{Wind turbine modeling using OpenFAST}
One critical requirement for incorporating a wind turbine solver into the DRL control framework is the ability to perform rapid evaluations, as the DRL training process requires a large number of simulations. To meet this need, we have chosen to employ an efficient Blade Element Momentum Theory (BEMT) solver. Specifically, we use OpenFAST \citep{OpenFAST}, a well-known open-source software for simulating wind turbine dynamics and acoustic predictions.

BEMT, is known for its efficiency and offers a simple yet accurate method for estimating the aerodynamic forces and energy generation of wind turbines. Its ability to perform rapid function evaluations is crucial for training and validating the agent within a reasonable timeframe. In the realm of BEMT, the wind turbine blade is segmented into smaller sections along its span. The aerodynamic forces exerted on each section are computed based on the local wind conditions and the airfoil's geometry. These local flow conditions, defined for every section and time step, encompass the wind's speed and direction, along with the turbulence intensity. Polar curves for each airfoil section are used to compute the aerodynamic forces (lift, drag and moment coefficients).
By integrating the forces along the span of the blades, we can derive the overall power and thrust generated by the wind turbine.

Additionally, OpenFAST includes an aeroacoustic module that enables the computation of noise levels generated by the wind turbine at specific observer locations. To determine the aerodynamic noise sources from wind turbine blades, various semi-empirical noise models are included \citep{zhu2005modeling} and we select the  Brooks Pope and Marcolini model. The sound pressure level (SPL) for each blade segment is calculated based on individual noise mechanisms. The cumulative effect of these mechanisms yields the noise source for each airfoil. Finally, the noise sources from all blade segments are combined as uncorrelated noise sources, contributing to the overall computation of the wind turbine's sound power level.
The essential aspect of this process is the precise identification and modeling of the various noise mechanisms associated with each blade section. These mechanisms can be categorized into two groups: turbulent inflow noise and airfoil self-noise. OpenFAST implements the turbulent inflow model presented by \cite{moriarty2004recent} and, among the airfoil self-noise models described by \cite{brooks1989airfoil}, we have specifically selected: turbulent boundary layer trailing edge noise and tip vortex formation noise.
\subsubsection{Validation of OpenFAST with a SWT2.3-93 wind turbine}
The onshore wind turbine selected for the study is based on the SWT2.3-93, which performs a rated power of 2.3 MW. This turbine has undergone extensive in field experimental testing and complete details on its geometry and benchmark data are available in the open access repository \href{https://zenodo.org/records/7323750}{Zenodo} (through the work of the European project \cite{zephyr}). The airfoil polar curves are available in the airfoil catalog compiled by \cite{bertagnolio2001wind}. More details can be found in \cite{matthew2012method}.

\begin{figure}[htbp]
  \centering
  \begin{subfigure}{0.45\textwidth}
    \centering
    \includegraphics[width=\linewidth]{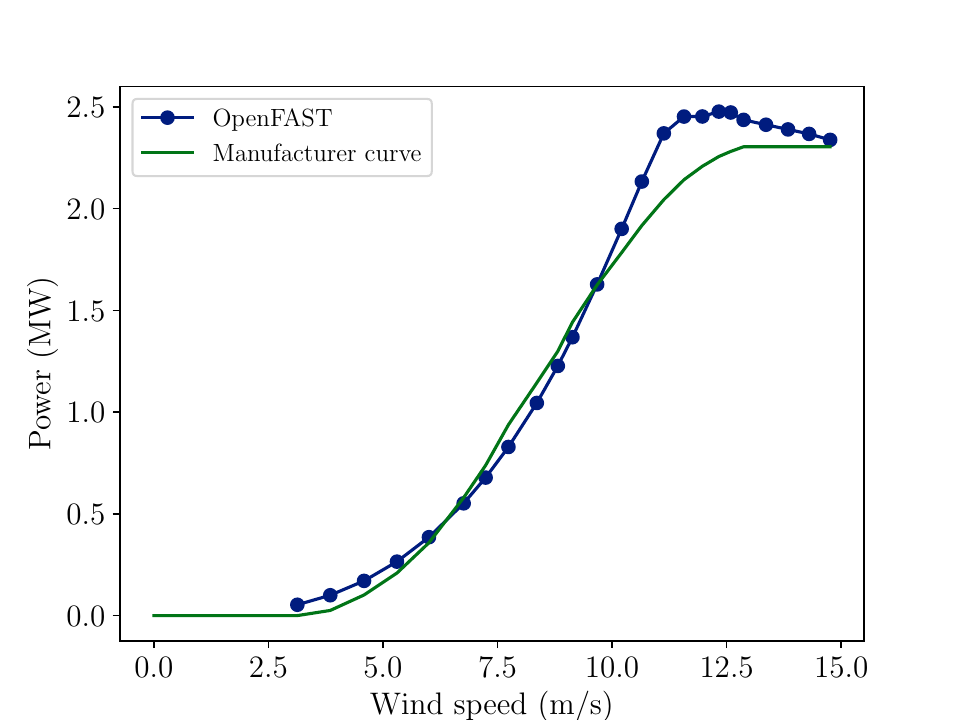}
    
    \caption{Power curve for the SWT2.3-93 wind turbine. The manufacturer curve is obtain from the \href{https://zenodo.org/records/7323750}{Zenodo} open access repository.}
    \label{fig:openfast_powercurve}
  \end{subfigure}
  \hfill
  \begin{subfigure}{0.45\textwidth}
    \centering
    \includegraphics[width=\linewidth]{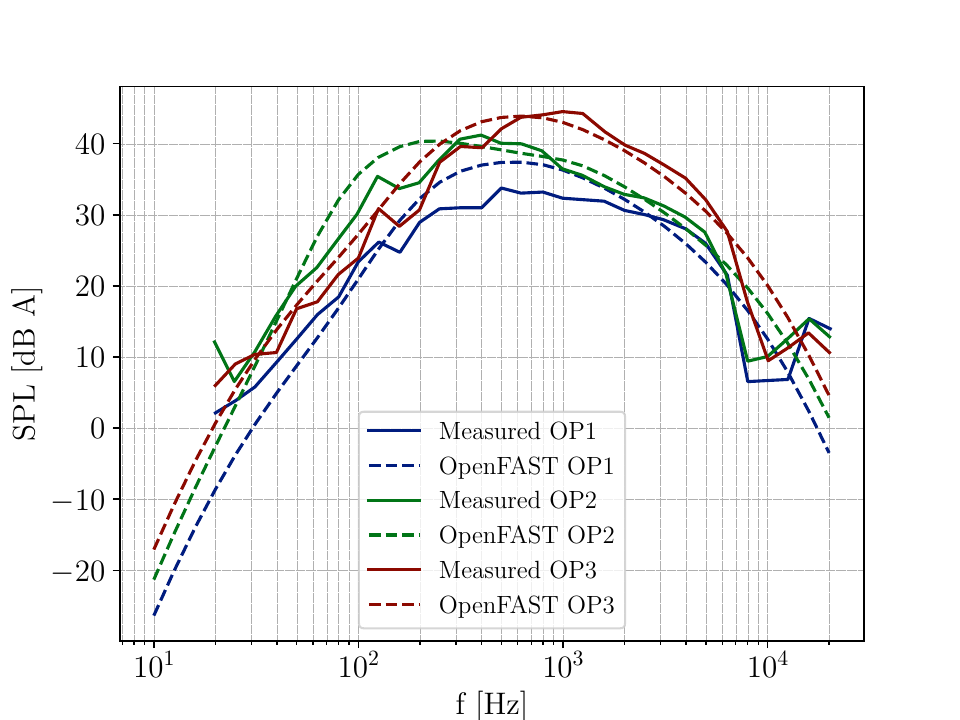}
    
    \caption{One-third octave SPL diagram for three different operational points. The measurements of the SPL spectrum are obtained from the \href{https://zenodo.org/records/7323750}{Zenodo} open access repository.}
    \label{fig:openfast_SPL}
  \end{subfigure}
  
  \caption{Comparison of \href{https://zenodo.org/records/7323750}{Zenodo} dataset benchmarks and OpenFAST modeling of the SWT2.3-93 wind turbine.}
  \label{fig:openfast_zenodo_comparison}
\end{figure}

All the open-source information enables us to create a SWT2.3-93 OpenFAST model. Additionally, the benchmark results from the Zenodo dataset can be utilized to validate the model. \Cref{fig:openfast_zenodo_comparison} presents the validation of both the power output and the sound pressure level (SPL) of the wind turbine. \Cref{fig:openfast_powercurve} compares the experimental power curve (from the Zenodo database) with that generated by the OpenFAST solver, showing good agreement. Meanwhile, \Cref{fig:openfast_SPL} shows the one-third octave SPL for frequencies ranging 10Hz to 10kHz, comparing the Zenodo dataset with results from OpenFAST. These comparisons cover three different operational conditions, detailed in \cref{tab:operational_conditions}. The Zenodo acoustic results are computed for an observer positioned 100 meters downstream from the wind turbine, at ground level. We observe good agreement with the experimental data for the three operating conditions. We conclude that OpenFAST, with the acoustic model of Brooks Pope and Marcolini, provides accurate predictions of power generation and acoustics, and is therefore a valid tool to perform multi-objective optimization.
\begin{table}[htbp]
    \centering
    \begin{tabular}{c|c c c }
    \hline
    \textbf{Operation point} & $\bm{U_{\infty}}$ \textbf{(m/s)} & $\bm\Omega$ \textbf{(rpm)} & $\bm\theta$ \textbf{(degrees)} \\ \hline\hline
    OP1 & 6 & 13 & 3 \\ 
    OP2 & 8 & 14 & $-2$ \\
    OP3 & 9.5 & 17 & 5 \\ \hline
    \end{tabular}
    \caption{Operational conditions studied on the Zenodo aeroacoustic dataset. The operation point is defined by the wind speed $U_\infty$, the rotational speed $\Omega$ and the blade pitch angle $\theta$.}
    \label{tab:operational_conditions}
\end{table}
\subsubsection{Sensitivity to control parameters}
\label{subsubsec:sensitivity}
The selected parameters to control the wind turbine power and noise include the rotational speed $\Omega$ and the blade pitch angle $\theta$. Before discussing the RL setup, it is crucial to illustrate the sensitivity of these parameters for the two performance metrics: the power coefficient and the overall sound pressure level.

\begin{figure}[htbp]
  \centering
  \begin{subfigure}{0.45\textwidth}
    \centering
    \includegraphics[width=\linewidth]{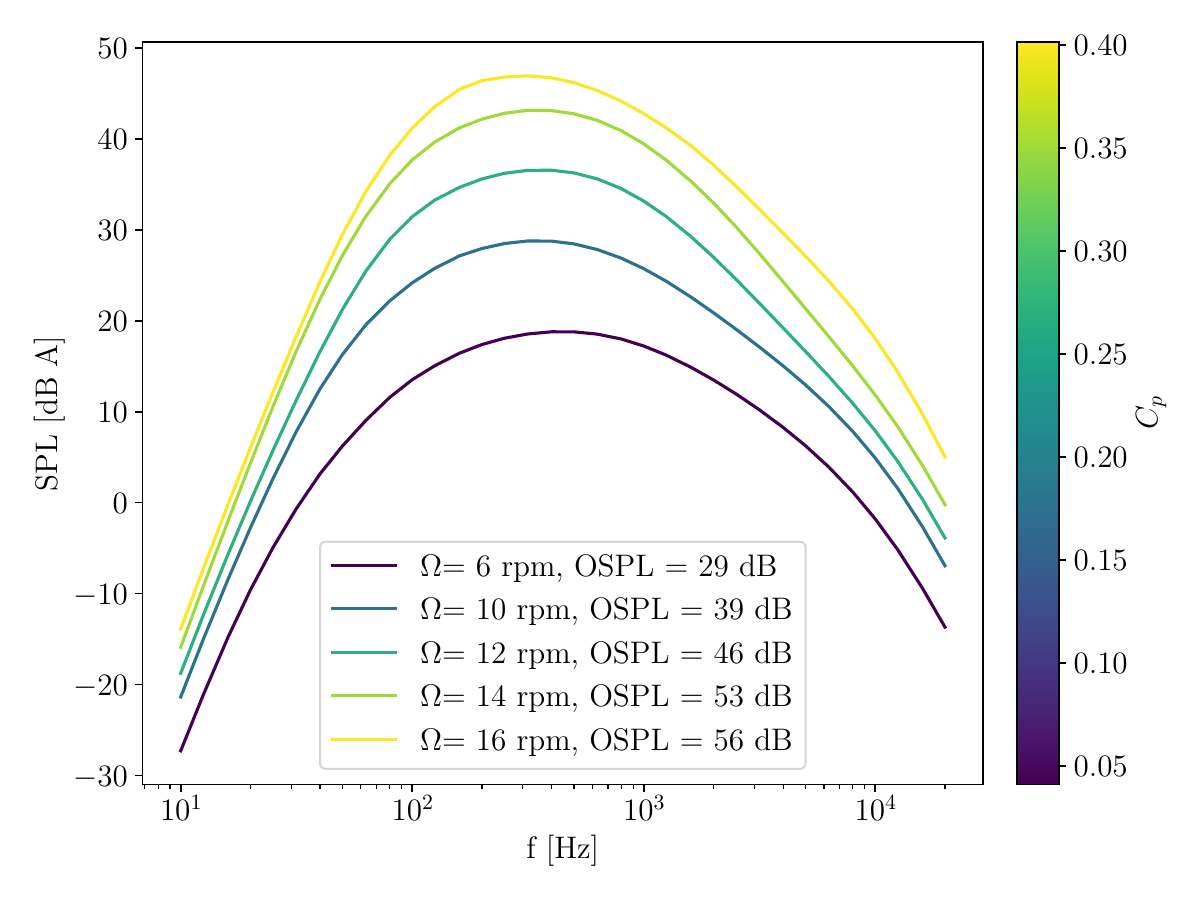}
    \captionsetup{justification=centering}
    \caption{Sensitivity analysis for rotational speed. \\ 
    Operational point : $U_\infty = 12$ m/s and $\theta = -1^\circ$.}
    \label{fig:sensitivity_rpm}
  \end{subfigure}
  \begin{subfigure}{0.45\textwidth}
    \centering 
    \includegraphics[width=\linewidth]{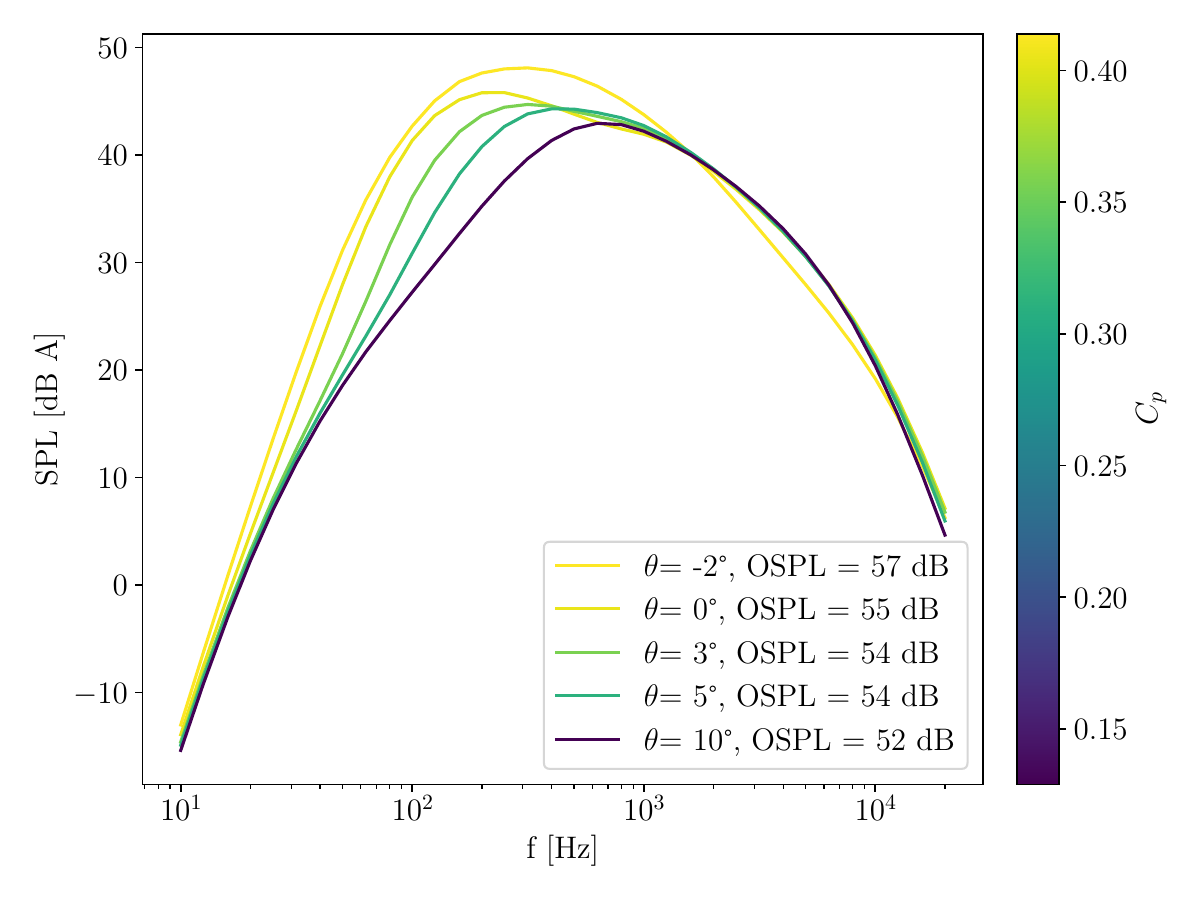}
    \captionsetup{justification=centering}
    \caption{Sensitivity analysis for pitch.\\ 
    Operational point : $U_\infty = 12$ m/s and $\Omega = 16.5$ rpm.}
    \label{fig:sensitivity_pitch}
  \end{subfigure}
  \caption{Sensitivity analyses for control parameters in relation to OSPL, power coefficient ($C_p$), and one-third octave SPL (dB A) spectra.}
  \label{fig:sensitivity_analysis}
\end{figure}

In \Cref{fig:sensitivity_analysis} the sensitivity analyses for both rotational speed and blade pitch angle are displayed for a single incoming wind speed ($U_\infty$). The values for one-third octave SPL, power coefficient and overall sound pressure level are shown for different operational conditions. \Cref{fig:sensitivity_rpm} depicts the influence of $\Omega$, increasing the rotational speeds makes both the sound pressure level and the power coefficient rise, highlighting the trade-off between maximizing power and minimizing noise. The entire SPL spectrum increases uniformly when increasing $\Omega$, due to a higher relative velocity in the blades and leading to a rise in SPL regardless of the noise mechanism. A similar analysis is presented in \Cref{fig:sensitivity_pitch} for the blade pitch angle. Although the general conclusion about the opposition between power and noise remains valid, the SPL spectrum behaves differently across frequencies. The pitch angle mainly affects trailing edge noise leading to changes at relatively low frequencies ranging from 10Hz to 1kHz. 

\subsection{Design of a reinforcement learning control}

Reinforcement Learning is a branch of machine learning that focuses on how agents should take actions in an environment to maximize cumulative reward. Unlike supervised learning, where the model learns from a labeled dataset, RL is driven by agent-environment interactions. 
The agent takes actions based on the current state of the environment and receives feedback in the form of rewards. The state represents the situation of the environment at a given time, while the actions are the possible moves the agent can make. The reward is the feedback indicating the immediate benefit or cost of an action, guiding the agent toward better actions over time.  In particular, in this work we use Q-learning RL, which is detailed in the next section.

\subsubsection{Reinforcement Learning for Multi Objective Control}
\label{subsubsec:RL}
Q-learning is a widely recognized reinforcement learning algorithm \citep{watkins1992q}. It is categorized under model-free RL algorithms, implying that it operates without the necessity for prior knowledge or explicit models that represent the dynamics of the system.
The fundamental component of Q-learning is the Q-value, which quantifies the anticipated cumulative reward for executing a specific action in a given state. The Q-value is updated iteratively via the Bellman equation, which formulates the optimal action-value function in terms of the maximum expected future reward.
During the learning process, the wind turbine interacts with the environment, transitions between states, and takes actions according to its current policy. The Q-learning algorithm employs an $\epsilon$-greedy exploration-exploitation trade-off to strike a balance between exploring new actions ($\epsilon$ times) and exploiting current knowledge ($1-\epsilon$ times) to maximize cumulative rewards. In RL the cumulative reward is computed taking into account that a reward received immediately is worth more than a reward received in the future, specifically, each time step the reward is discounted by $\gamma$, the discount rate. 

\begin{figure}[htbp]
    \centering
    \begin{tikzpicture}[scale=1]
    
    % Colors
    \definecolor{darkblue}{RGB}{0,28, 127} 
    \definecolor{darkgreen}{RGB}{1, 117, 23} 
    \definecolor{darkred}{RGB}{140, 9, 0}
    % Nodes
    \node[draw, fill=darkgreen!20, rounded corners, minimum width=7cm, minimum height=5cm, anchor=north west] (AgentBox) at (0,0) {};
    \node[draw, fill=darkblue!20, rounded corners, minimum width=3cm, minimum height=5cm, anchor=north east] (environment) at (12,0) {};
    \node[draw, fill=darkblue!20, rounded corners, minimum width=1.5cm, minimum height=1cm, anchor=north west] (state) at (0.5,-2) {State};
    
    % Neural Network
    \foreach \i in {1,...,5}
    {
        \foreach \j in {1,2}
        {
            \node[circle, fill=gray!30, draw=black, minimum size=0.5cm] (layer-\j-n-\i) at (2+\j,0.75*\i-5) {};
        }
    }
    
    \foreach \i in {1,...,4}
    {
            \node[circle, fill=gray!30, draw=black, minimum size=0.5cm] (layer-3-n-\i) at (5,0.75*\i-4.5) {};
    }

    % Connections
    \foreach \i in {1,...,5}
    {
        \draw[->] (state) -- (layer-1-n-\i); 
        \foreach \k in {1,...,4}
        {
            \draw[->] (layer-2-n-\i) -- (layer-3-n-\k);
        }
    }
    
            \foreach \i in {1,...,5}
    {
        \foreach \k in {1,...,5}
        {
            \draw[->] (layer-1-n-\i) -- (layer-2-n-\k);
        }
    }
    
    \node[draw, fill=blue!30, minimum width=0.3cm, minimum height=0.3cm, anchor=north west] (output1) at (5.5,0.75*1-4.35) {};
    \node[draw, fill=blue!30, minimum width=1cm, minimum height=0.3cm, anchor=north west] (output1) at (5.5,0.75*2-4.35) {};
    \node[draw, fill=blue!30, minimum width=0.6cm, minimum height=0.3cm, anchor=north west] (output1) at (5.5,0.75*3-4.35) {};
    \node[draw, fill=blue!30, minimum width=0.1cm, minimum height=0.3cm, anchor=north west] (output1) at (5.5,0.75*4-4.35) {};
    \node[below left=3cm and 2.1cm of AgentBox.center] (aux1) {};
    \node[below=3cm of environment.center] (aux2) {};
    \node[above=0.75cm of environment.north] (aux3) {}; 
    \node[above=0.75cm of AgentBox.north] (aux4) {};
    \node[below=0.75cm of AgentBox.north] (aux5) {};
    % Arrows
    \draw[-] (5.5,-1) -- (5.5,-4.25);
    \draw[->] (6.5,-2.5) -- (environment.west) node[midway, above] (action_node){Action};
    \draw[-] (aux1.center) -- (aux2.center) node[midway,above] (observation_node) {Observation};  
    \draw[->] (aux1.center) --(state.south); 
    \draw[-] (environment.south) -- (aux2.center);
    \draw[-] (aux3.center) -- (aux4.center) node[midway,below] (reward_node) {Reward}; 
    \draw[-] (environment.north) -- (aux3.center);  
    \draw[->] (aux4.center) -- (aux5.center) ; 
    \node[below = 3mm of observation_node.center] {$s_t$(wind conditions, control variables)};
    \node[above=3mm of reward_node.center] {$r_t$(Power, SPL)};
    \node[below=3mm of action_node.center] {$a_t$}; 
    % Labels
    \node[] at (6.1,-1) {$Q(s,a)$};
    \node[draw, fill=white,rounded corners,below right=3mm and 5mm of AgentBox.north west] {Agent};
    \node[draw, fill=white,rounded corners, below=5mm of environment.north] {Environment};
    \node[below = 20mm of environment.north] {OpenFAST};
    \node[below = 25mm of environment.north] {+};
    \node[below = 30mm of environment.north] {Wind conditions};
\end{tikzpicture}
    \caption{Flow diagram of the reinforcement learning control methodology.}
    \label{fig:RL_figure}
\end{figure}
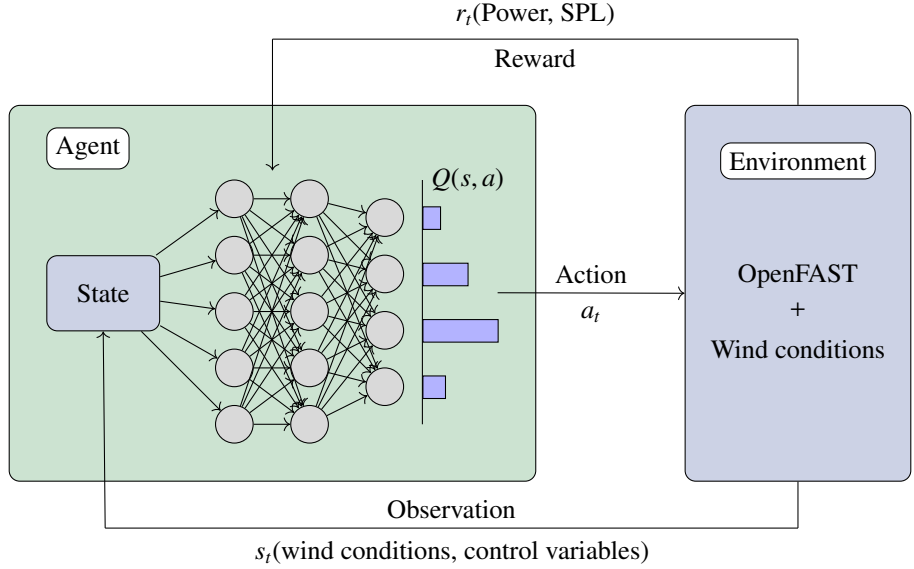

Initially, the Q-values are arbitrarily initialized. As the wind turbine explores the environment and receives feedback in the form of rewards, the Q-values are updated using the temporal difference error. The temporal difference error represents the discrepancy between the observed reward and the predicted reward based on the Q-values.
Through repeated iterations, the Q-learning algorithm gradually converges to an optimal policy. In this state, the wind turbine learns the best actions to take in different states, thereby maximizing power generation while minimizing noise. In our case, this agent-environment interactions for the wind turbine control are illustrated on \Cref{fig:RL_figure}.

Deep Q-Network (DQN) is a variant of Q-learning that employs a deep neural network to estimate Q-values \citep{mnih2013playing}. It replaces the traditional lookup table with a neural network, enabling generalizations across states to handle large state spaces efficiently. In this study, a Double Deep Q-Learning (DDQN) is employed. DDQN is an extension of DQN that uses two neural networks: the primary network and the target network. The primary network select the action and the target network evaluates its Q-value. This way of decoupling the action selection and evaluation addresses the overestimation of Q-values, often observed in DQN algorithms due to the maximization bias, \citep{van2016deep}. The weights of the primary network are obtained by minimizing the following loss function:
\begin{equation}
\label{eq:loss_DDQN}
    \mathcal L(\phi) = \mathbb{E}_{(s, a, r, s')} \left[ \left( r + \gamma Q_{\phi'} \left( s', \arg\max_a Q_{\phi}(s', a) \right) - Q_{\phi}(s, a) \right)^2 \right],
\end{equation}
where $r$ is the reward, $s$ is the state of the environment, $a$ denotes a possible action that the agent can take, $Q(s,a)$ is the Q-Function and $\phi$ and $\phi'$ are the set of weights of the primary and target network, respectively. The loss function $\mathcal L(\phi)$ quantifies the residual of the Bellman equation, which defines formally the optimal values of the Q-values, \citep{sutton1998rli}. The set of weights from the target network , $\phi'$, is updated using a soft update rule to enhance the stability of the learning process, \citep{lillicrap2015continuous}.
\begin{equation}
\label{eq:soft_update}
    \phi' \leftarrow \tau \phi + (1 - \tau) \phi'.
\end{equation}

To train the DDQN, an experience replay buffer is utilized. During the training phase, the agent interacts with the environment and stores the experiences (state, action, reward, next state) in the replay buffer. Subsequently, random batches of experiences are sampled from the replay buffer to train the network and update its weights. This process helps to break the correlation between consecutive samples and improves stability during the learning process.

An additional consideration in solving this reinforcement learning problem is the need to balance maximizing power output with minimizing noise impact. These objectives are inherently conflicting, placing this problem within the Multi-Objective Reinforcement Learning (MORL) framework. MORL extends traditional reinforcement learning to handle problems involving multiple, often conflicting, objectives.

Various strategies exist for addressing MORL problems. One of the simplest methods is to define the reward using a scalarized function that combines the rewards for each objective into a single global reward, thus transforming the problem into a single-objective reinforcement learning task \citep{van2013scalarized}. Another approach involves Pareto optimization, which aims to find a set of optimal policies that lie on the Pareto Front, where no other policy is superior in all objectives \citep{van2014multi}. There are already methods that apply these MORL approaches using deep learning implementations \citep{mossalam2016multi}. In this work, a scalarized method is adopted to define a reward that balances the two objectives of maximizing power and minimizing noise.

\subsubsection{State-action structure}
The state of the agent must include all the necessary information about the environment to enable the agent to take the best possible action. If the state lacks relevant information, the agent may not be able to achieve optimal performance. The state of the agent is defined by the incoming wind conditions, specifically the wind speed, $U_\infty$, along with the control variables of the wind turbine, which are the rotational speed, $\Omega$, and the blade pitch angle, $\theta$. To fit within the DDQN framework, the state space $\mathcal S$ needs to be bounded. Some variables (rotational speed and pitch) are bounded by mechanical/structural limitations, whereas the wind speed is bounded by physical range. Note that these can be tuned for specific wind turbines and geographic sites.  We include an additional constraint on the tip speed ratio $\lambda = \frac{\Omega R}{U_\infty}$, with the blade radius $R=46.5$ m, to ensure the correct behavior of the BEMT solver. The specific values of all the constraints are outlined below:   
\begin{itemize}
    \item $U_\infty\in$ [4,16] m/s,
    \item $\Omega\in [6,18]$ rpm,
    \item $\theta\in [-5,10]$ degrees,
    \item $\lambda \in [3,12]$.
\end{itemize}
The actions available to the agent involve either increasing or decreasing the control variables. Since Q-learning is defined for a discrete action space $\mathcal A$, the control variables can only be adjusted by fixed increments. Five distinct actions are defined: two for each control variable (one for increasing and one for decreasing), and one for maintaining the current state (doing nothing). The specific fixed increment for each possible action is determined based on the sensitivity analysis detailed in \Cref{subsubsec:sensitivity}. Since the rotational speed is more sensitive to both power generation and sound pressure level compared to the pitch angle, the incremental adjustments for each variable has been designed so that their corresponding actions have effects of the same magnitude. The actions that the agent can take are specified as follows:
\begin{itemize}
    \item $a_1$: increase $\Omega$ by 0.5 rpm,
    \item $a_2$: decrease $\Omega$ by 0.5 rpm, 
    \item $a_3$: increase $\theta$ by 1 degree,
    \item $a_4$: decrease $\theta$ by 1 degree, 
    \item $a_5$: do nothing.
\end{itemize}

It is important to note that the transition between states is not deterministic \textit{a priori}. Although we can freely adjust the control variables, the wind conditions depend on the environment and are beyond our control. This motivates the use of a model-free reinforcement learning method, as model-based approaches only guarantee convergence if the transition function between states is known.

\subsubsection{Reward definition}
The reward function is key when defining the RL algorithm, as it is the only feedback to quantify how successful are the actions taken by the agent. Therefore, the reward must be carefully crafted for each specific problem to learn an appropriate policy.
As mentioned in \Cref{subsubsec:RL}, this is a multi-objective optimization problem (or MORL), requiring a specific strategy to address the two conflicting objectives: maximizing power extraction while minimizing noise generation. In this work, we choose to blend the two objectives through a linear function, to define the overall reward. The reward can be expressed as follows:
\begin{equation}
\label{eq:linear_reward}
    r =  r_{\text{PW}} + r_\SPL,
\end{equation}
where  $r_{\text{PW}}$ denotes the reward associated to the power objective and $r_\SPL$ the one related to the SPL one. 

\begin{figure*}[htbp]
  \centering
  \begin{tabular}{cc}
    \begin{minipage}{0.4\textwidth}
      \centering
      \includegraphics[width=\textwidth]{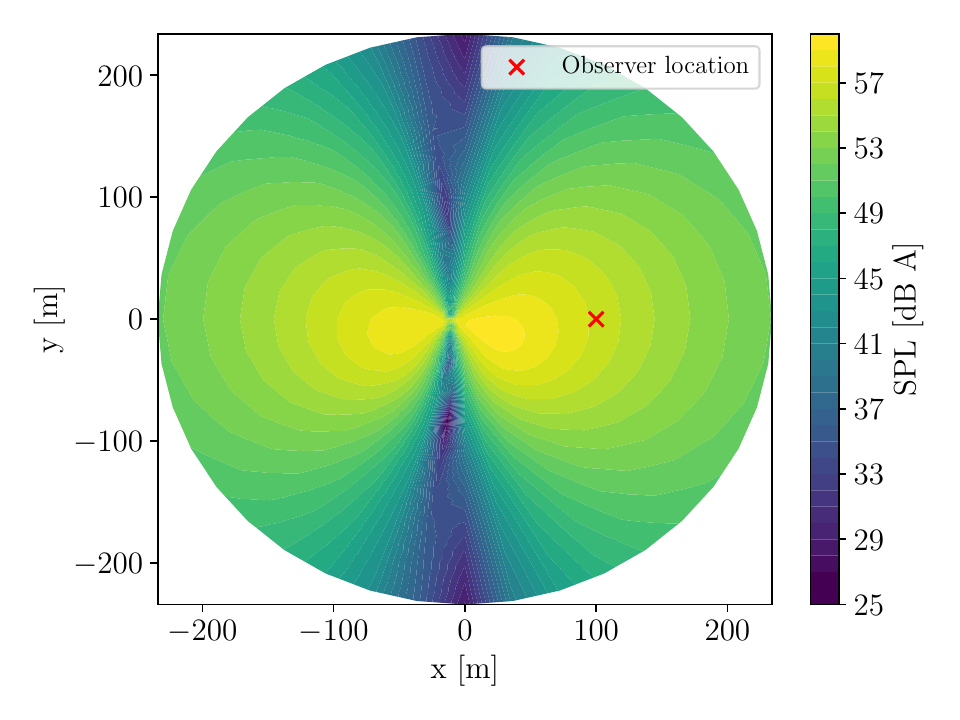}
      \caption{Directivity map of Sound Pressure Level (SPL) generated by OpenFAST. The wind direction is oriented along the positive $x$ axis (left to right), perpendicular to the wind turbine rotor. The operational conditions are  \\ 
      $U_\infty = 12$ m/s, $\Omega = 16.5$ rpm and $\theta = -1$ degree.\\ 
      The observer location '\textcolor{red}{$\times$}' is situated 100 m downwind.}
      \label{fig:directivity_map}
    \end{minipage} &
    \begin{minipage}{0.4\textwidth}
      \centering
       \begin{tikzpicture}
        \def\SPLthreshold{2}
        \def\Deltadb{1}
        
        \begin{axis}[
            axis lines=middle,
            xlabel style={below},
            xlabel=$\SPL$,
            ylabel=$r_\SPL$,
            xtick=\empty,
            ytick={-1},
            domain=0:4,
            samples=100,
            width=5cm,
            height=5cm,
            legend pos=north west
        ]
       
        \addplot [
            domain=-1:\SPLthreshold,
            samples=2,
            very thick,
            blue
        ] {0};
        \addplot [
            domain=\SPLthreshold:4,
            samples=2,
            very thick,
            blue
        ] {\SPLthreshold-x};
        \addplot[dashed, gray] coordinates {(\SPLthreshold,0) (\SPLthreshold,\SPLthreshold)};
        \node at (axis cs:\SPLthreshold, +0.3) [anchor=east] {$\text{\small{SPL}}_{\text{thr}}$};
        \addplot[dashed, gray] coordinates {(\SPLthreshold+\Deltadb,-1) (\SPLthreshold+\Deltadb,\SPLthreshold)};
        \addplot[dashed, gray] coordinates {(0,-1) (3,-1)};
        \draw[<->] (\SPLthreshold, 1) -- (\SPLthreshold +\Deltadb, 1);
        \node at (axis cs:\SPLthreshold + 0.5, 1.6) [anchor=north] {\small{$\Delta$dB}};
        \end{axis}
    \end{tikzpicture}
      \caption{Reward noise component.}
      \label{fig:relu_reward}
    \end{minipage} 
  \end{tabular}
\end{figure*}

As we already discussed in our previous work \cite{SOLER2024123502}, the reward power component should encourage the agent to obtain the highest energy generation possible, regardless of the wind conditions. To achieve this, we use the power coefficient $C_p$ of the wind turbine. We set this reward to increase linearly from 0 to 1, with 1 corresponding to the maximum possible value of the power coefficient within the state space, $C_{p,\text{nom}}$. Therefore, the reward power component reads,   
\begin{equation}
\label{eq:power_reward}
    r_{\text{PW}} = \frac{C_p}{C_{p, {\text{nom}}}}.
\end{equation}
The reward term related to sound generation, $r_\SPL$, is highly dependent on the specific problem being modeled. First, we need to select the observer locations where the SPL is computed, typically in critical areas where noise mitigation is a priority. In this work, we decide to set one observer 100 m downstream the wind turbine, see \Cref{fig:directivity_map}. Next, we decide how to penalize the sound generation (SPL) in the reward function. We opt to use a ReLU activation function that begins penalizing once the SPL exceeds a certain threshold, $\SPL_\text{thr}$. Below this threshold, the agent focuses solely on maximizing power. Additionally, we define a $\Delta$dB value that specifies how much the SPL threshold can be exceeded before the reward becomes $-1$. Beyond this point, no matter how much power the agent generates, the total reward will be negative. Therefore, 
$\SPL_\text{thr}+\Delta$dB serves as an effective noise limit. For this specific application, we defined $\SPL_\text{thr} = 45$ dB and $\Delta$dB = 5 dB, but note that these values can be adapted to specific sites or regulations. The reward noise component can be seen in \Cref{fig:relu_reward} and reads as follows:
\begin{equation}
\label{eq:SPL_reward}
    r_\SPL = -\text{ReLU}\left(\frac{\SPL-\SPL_{\text{thr}}}{\Delta \text{dB}}\right).
\end{equation}

In addition, we need to include the bounds of $\mathcal S$ in the reward. To make the agent learn the limits, it receives punishments whenever it performs a forbidden action, that is, an action that leads to a state $s_{t+1}\notin\mathcal S$. In such cases, the agent receives a negative reward with a value of $r = -3$ and the action is revoked so that the control variables remain the same. The punishment is set to $-3$ to differentiate it from the possible negative reward of $r_\SPL$. This distinction is made because exceeding the $\mathcal S$ limits is considered worse than generating noise above the threshold.
Finally, the reward function for the agent is the following
\begin{equation}
\label{eq:complete_reward}
    r(s_t,a_t,s_{t+1}) =
    \begin{cases}
        \displaystyle\frac{C_p(s_{t+1})}{C_{p,nom}} -\text{ReLU}\left(\frac{\SPL(s_{t+1})-\SPL_{\text{thr}}}{\Delta \text{dB}}\right),\\
         \qquad -3 \qquad \text{if } s_{t+1}\notin\mathcal S.\\
    \end{cases}
\end{equation}

\subsubsection{Neural Network architecture}
When using DQN, neural networks (NN) are employed to approximate the Q-Function. Typically, the NN is designed to approximate the Q-Vectors, $\bm  q(s)$, which represent the Q-Values in the state $s$ for all possible actions. That is, $\bm q(s)_i = Q(s,a_i)$. This approach is used because $\mathcal A$ is a discrete space, and encoding these discrete actions as inputs can be problematic; it is more convenient to create a mapping between real subspaces.  The neural network map is defined as $\bm q_\phi(\bm s): \mathcal S\subset\mathbb R^3 \to \mathbb R^5$, where $\phi$ denotes all the NN weights, the output space dimension is $|\mathcal A|=5$ and $\bm s$ denotes the state vector, which is $\bm s = [U_\infty,\Omega,\theta]^T$.

\begin{figure}[htbp]
    \centering
    \begin{tikzpicture}[scale=0.75,node distance=1.5cm]
    % Colores acordes con las gráficas del paper 
    \definecolor{darkblue}{RGB}{0,28, 127} 
    \definecolor{darkgreen}{RGB}{1, 117, 23} 
    \definecolor{darkred}{RGB}{140, 9, 0}
   
    % Capa de entrada
    \node[circle, fill=darkgreen!20,draw=darkgreen!60, line width=1.5pt, minimum size=0.5cm] (input-1) at (0,-2) {};
    \node[circle, fill=darkgreen!20,draw=darkgreen!60, line width=1.5pt, minimum size=0.5cm] (input-2) at (0,-3) {};
    \node[circle, fill=darkgreen!20,draw=darkgreen!60, line width=1.5pt, minimum size=0.5cm] (input-3) at (0,-5) {};
    
    % Primera Capa oculta
    \node[circle, fill=darkblue!20,draw=darkblue!60, line width=1.5pt, minimum size=0.5cm] (hidden-1) at (2,-1.5) {\tiny{}};
    \node[circle, fill=darkblue!20,draw=darkblue!60, line width=1.5pt, minimum size=0.5cm] (hidden-2) at (2,-3) {\tiny{}};
    \node[circle, fill=darkblue!20,draw=darkblue!60, line width=1.5pt, minimum size=0.5cm] (hidden-3) at (2,-5.5) {\tiny{}};
    
    % Puntos suspensivos en la capa inicial
    \node at (0,-3.85) {$\vdots$};
    
    % Segunda capa oculta 
    \node[circle, fill=darkblue!20, draw=darkblue!60, line width=1.5pt, minimum size=0.5cm] (hidden2-1) at (4,-1.5) {\tiny{}};
    \node[circle, fill=darkblue!20, draw=darkblue!60, line width=1.5pt, minimum size=0.5cm] (hidden2-2) at (4,-3) {\tiny{}};
    \node[circle, fill=darkblue!20, draw=darkblue!60, line width=1.5pt, minimum size=0.5cm] (hidden2-3) at (4,-5.5) {\tiny{}};

    % Tercera capa oculta 
    % \node[circle, fill=darkblue!20, draw=darkblue!60, line width=1.5pt, minimum size=0.5cm] (hidden3-1) at (6,-1.5) {\tiny{}};
    % \node[circle, fill=darkblue!20, draw=darkblue!60, line width=1.5pt, minimum size=0.5cm] (hidden3-2) at (6,-3) {\tiny{}};
    % \node[circle, fill=darkblue!20, draw=darkblue!60, line width=1.5pt, minimum size=0.5cm] (hidden3-3) at (6,-5.5) {\tiny{}};
    
    % Capa de salida  
    \node[circle, fill=darkred!20, draw=darkred!60, line width=1.5pt, minimum size=0.5cm] (hidden4-1) at (6,-2) {\tiny{}};
    \node[circle, fill=darkred!20, draw=darkred!60, line width=1.5pt, minimum size=0.5cm] (hidden4-2) at (6,-3) {\tiny{}};
    \node[circle, fill=darkred!20, draw=darkred!60, line width=1.5pt, minimum size=0.5cm] (hidden4-3) at (6,-5) {\tiny{}};
    
    % Puntos suspensivos en las capas ocultas
    \node at (2,-3.85) {$\vdots$};
    \node at (4,-3.85) {$\vdots$};
    \node at (6,-3.85) {$\vdots$};
    % \node at (8,-3.85) {$\vdots$};

    % Conexiones
    \foreach \i in {1,2,3}
    \foreach \j in {1,2,3}
    \draw[->] (input-\i) -- (hidden-\j);
    
    \foreach \i in {1,2,3}
    \foreach \j in {1,2,3}
    \draw[->] (hidden-\i) -- (hidden2-\j);
    
    \foreach \i in {1,2,3}
    \foreach \j in {1,2,3}
    \draw[->] (hidden2-\i) -- (hidden4-\j);

    % \foreach \i in {1,2,3}
    % \foreach \j in {1,2,3}
    % \draw[->] (hidden3-\i) -- (hidden4-\j);
    
    % Etiquetas
    \node at (0,-6) {\footnotesize{$\bm s$}};
    % \node at (2,-6.5) {\footnotesize{128 neurons}};
    \node at (2,-6.5) {\footnotesize{128 neurons}};
    \node at (4,-6.5) {\footnotesize{64 neurons}};
    \node at (6,-6) {\footnotesize{$\bm q_\phi(\bm s)$}};
    \node at (7.5,-2) {\footnotesize{$Q(s,a_1)$}};
    \node at (7.5,-3) {\footnotesize{$Q(s,a_2)$}};
    \node at (7.5,-3.85) {$\vdots$};
    \node at (7.5,-5) {\footnotesize{$Q(s,a_5)$}};
\end{tikzpicture}
    %\captionsetup{justification=centering}
    \caption{Q-Network architecture.}
    \label{fig:NN_architecure}
\end{figure}
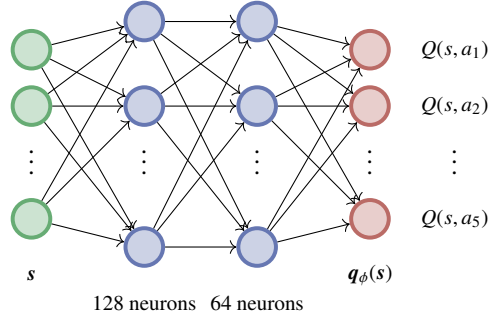

The neural network architecture employs a Multi-Layer Perceptron structure, consisting of two dense hidden layers with Rectified Linear Unit (ReLU) activation functions. The final layer, uses a linear activation function instead of ReLU. This allows the Q-values to take on any sign, rather than being restricted to positive numbers. The number of dense layers and their sizes were determined through extensive trial and error. Ultimately, two layers with 128 and 64 neurons, were found to be sufficient to accurately represent the Q-Function. The architecture of the Q-Network used to train the DDQN agent is shown in \Cref{fig:NN_architecure}.

\subsubsection{Training of the Wind Turbine DDQN Agent}
The main ideas of Q-Learning have already been explained in \Cref{subsubsec:RL}. However, here the specific details of the DDQN training to design the wind turbine controller are included. During the training phase, the agent faces random steady wind
conditions during short episodes of 20 time steps. This allows the agent to adapt to virtually any wind, even if the wind speed changes faster than 20 time steps. This adaptability is achieved because experiences are stored in the replay buffer, and batches are selected randomly. Consequently, the specific temporal evolution of states during the agent's experience is not critical, provided that the stored transitions comprehensively represent all actual transitions in the system.

\begin{algorithm}
\begin{algorithmic}[1]
    \STATE Initialize primary network $Q_{\phi}$ with random weights $\phi$
    \STATE Initialize target network $Q_{\phi'}$ with weights $\phi' = \phi$
    \STATE Initialize replay buffer $\mathcal R$
    \STATE Set hyperparameters: $\alpha$ (learning rate), $\gamma$ (discount factor), $\epsilon$ (exploration probability), $\tau$ (soft update parameter), $N$ (number of training iterations), $n$ (number of steps taken every training step, $m$ (period of update of the target network). 
    \STATE Initialize random state $s$
    \FOR{iter=1 \TO N}
        \FOR{n steps}
            \STATE Select action $a$ using $\epsilon$-greedy policy from $Q_{\phi}$
            \STATE Take action $a$, compute reward $r$ and new state $s'$ using the wind turbine solver 
            \STATE Store transition $(s, a, r, s')$ in replay buffer $\mathcal R$
            \IF{last step of the episode}
                \STATE Initialize random state $s$
            \ELSE
                \STATE Update current state: $s \leftarrow s'$
            \ENDIF 
        \ENDFOR 
            \STATE Sample random mini-batch of transitions $(s_j, a_j, r_j, s_j')$ from $\mathcal R$
            
            \STATE Estimate loss function from \cref{eq:loss_DDQN} with the mini-batch transitions
            
            \STATE Update the primary set of weights $\phi$ using a gradient optimizer. \\ 
            $\phi \leftarrow \phi - \alpha\nabla_\phi\mathcal L$

            \STATE \textbf{each} m iterations \textbf{do} Soft update of the target network weights $\phi'$ using \cref{eq:soft_update} rule.
    \ENDFOR
\end{algorithmic}
\caption{Double Deep Q-Learning Algorithm}
\label{alg:ddqn}
\end{algorithm}

\begin{figure*}[htbp]
    \begin{minipage}{0.4\textwidth}
  \includegraphics[width=7cm]{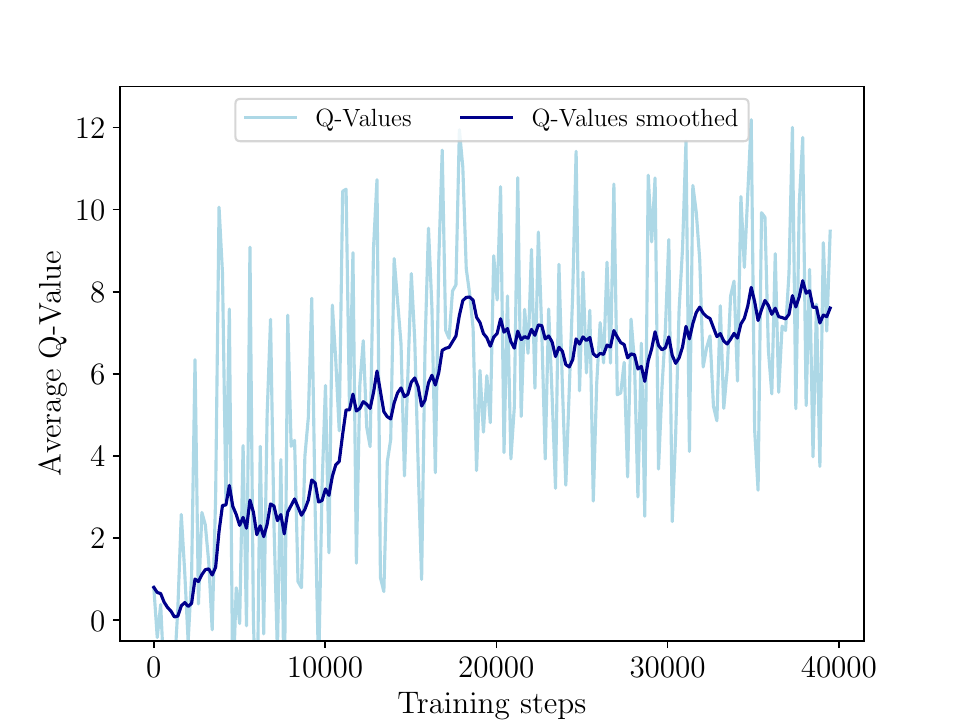}
   \caption{Average Q-Values taken during the training process of the DDQN agent.}
    \label{fig:trainQV}
\end{minipage}\hfill
\begin{minipage}{0.55\textwidth}
    \centering
    \begin{tabular}{ll}
        \hline
        \textbf{Parameter} & \textbf{Value} \\
        \hline
        Environment interactions & 200k \\
        Steps of the environment per training iteration & 5 \\
        Maximum capacity of the replay buffer & 50k \\
        Batch size & 64 \\
        Learning rate & 5e-4 \\
        Discount factor & 0.95 \\
        Epsilon value for the epsilon greedy policy & 0.50 \\
        Tau soft update parameter & 0.1 \\
        Period of update of the target network & 20 \\
        \hline 
    \end{tabular}
   \captionof{table}{Training parameters used for the DDQN agent.}
   \label{tab:hyperparameters}
\end{minipage}
\end{figure*}

The Double Deep Q-Network (DDQN) agent is trained using the hyperparameters listed in \Cref{tab:hyperparameters}. To illustrate the influence of each parameter on the training process, a pseudocode for the DDQN training is presented in \Cref{alg:ddqn}. The effectiveness of the learning progress during training is assessed by displaying the Q-values of the state-action pairs encountered by the agent, as shown in \Cref{fig:trainQV}. As the agent learns, the Q-values of the actions taken at each state are expected to increase, as depicted in the figure.
For the implementation, we utilized OpenAI Gym \citep{Gymnasium} to create the environment, serving as a bridge between the reinforcement learning formulation and the OpenFAST wind turbine solver. TF-Agents \citep{TFAgents} was employed to develop the agent and manage the entire training process. All neural networks were constructed using the Keras API \citep{keras}, and the Adam optimizer was used for training \citep{kingma2014adam}.

\section{Results}
\label{sec:results}
The agent's performance is assessed under various wind conditions. First, we validate the operational point that the agent reaches under constant wind conditions, assessing its optimality via a Pareto front. Next, we evaluate the agent's ability to adapt to turbulent wind conditions, comparing its control strategy against a classic controller. Finally, we estimate the agent's annual energy production and compare it against a classic control strategy designed to maximize energy extraction.

\subsection{Steady wind validation}
The simplest test for evaluating the agent is to assess its performance under steady wind conditions. In this scenario, with unchanging wind conditions, the agent should identify, reach  and maintain the state that maximizes the cumulative reward. This optimum state does not depend on the initial conditions of pitch and rotor speed, as the wind conditions are steady. However, it is important to note that the agent's actions are discrete, limiting its ability to reach every possible state.

\begin{figure*}[htbp]
    \begin{minipage}{0.5\textwidth}
      \centering
    \includegraphics[width=8cm]{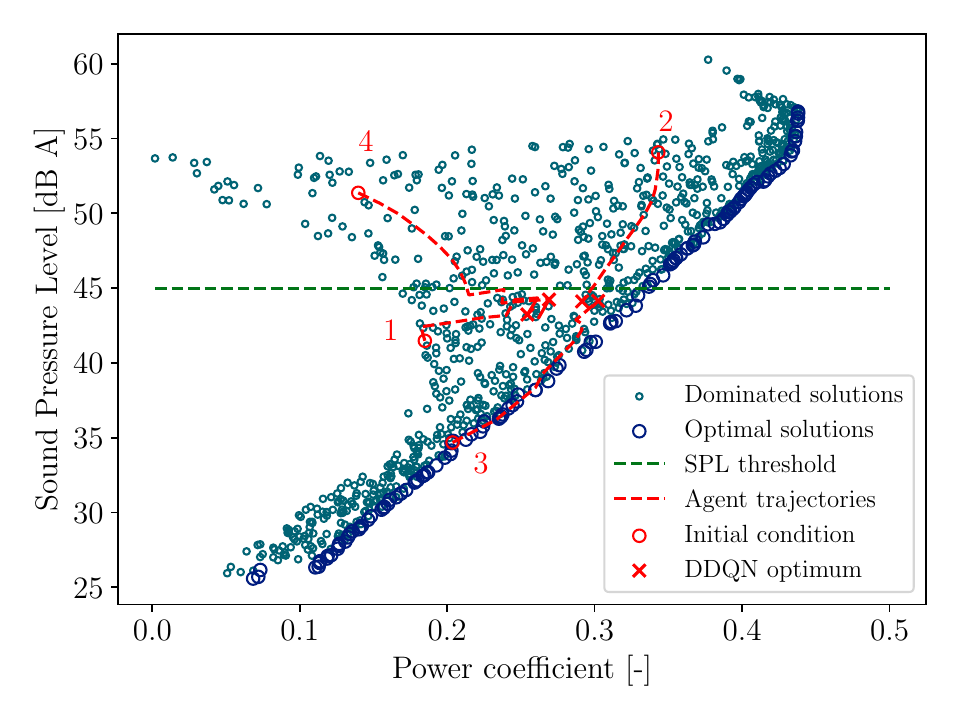}
    %\captionsetup{justification=centering}
    \caption{Pareto diagram at a wind speed of 10 m/s, showing the power coefficient $C_p$ and sound pressure level SPL (dB A) computed for 1000 random states within the state space. Different agent trajectories are displayed on the Pareto diagram. The numbering of the initial conditions correspond to \Cref{tab:initial_conditions}.}
    \label{fig:Pareto}
\end{minipage}\hfill
\begin{minipage}{0.45\textwidth}
    \centering
    \begin{tabular}{c c c c c}
        \hline
        \textbf{Case} & \(\mathbf{\Omega_0}\) \textbf{(rpm)} & \(\mathbf{\Omega_f}\) \textbf{(rpm)} & \(\mathbf{\theta_0}\) \textbf{(°)} & \(\mathbf{\theta_f}\) \textbf{(°)} \\
        \hline 
        Case 1 & 9.61 & 11.11 & 9.94 & 5.93 \\
        Case 2 & 17.98 & 10.98 & 3.87 & 3.87 \\
        Case 3 & 8.43 & 10.93 & -1.25 & 2.75 \\
        Case 4 & 16.54 & 11.04 & 8.37 & 5.37 \\
        \hline
    \end{tabular}
   \captionof{table}{Four different cases evaluated under steady wind speed of 10 m/s. The initial condition is defined by the initial control variables $\Omega_0$ and $\theta_0$ and the agent get to the optimum defined by the control variables $\Omega_f$ and $\theta_f$.}
   \label{tab:initial_conditions}
\end{minipage}
\end{figure*}

To validate the performance and robustness of the agent, a Pareto diagram is used. The agent is tested for different initial conditions (with the same steady wind speed) to determine if it can consistently reach optimal states (at the pareto front), regardless of the initial state. To illustrate the agent's trajectory (sequence of state-action-reward) for each initial condition, these trajectories are displayed on a power coefficient - sound pressure level diagram, along with values for 1000 random states from $\mathcal S$. 
\Cref{fig:Pareto} presents the Pareto diagram, illustrating the agent's trajectory from four different initial conditions. It is clear that, regardless of the initial condition, the agent successfully achieves high power outputs up to the maximum permissible decibel level. Furthermore, the agent demonstrates robust performance by consistently avoiding the maximum limit of SPL (dB A) while remaining close to the limit to maximize power. It can be seen that the RL does not always reach the same final state, but that the optima are relatively close to each other. This suggests the existence of local optimum. In addition, the  discrete nature of the Q-learning actions, may not allow the agent to reach certain optima, since not all states are reachable from an initial state. Despite these issues, the agent consistently avoids acoustic penalties and achieves high power outputs, with power coefficients ranging from 0.26 to 0.30.

For completness, \Cref{tab:initial_conditions} shows the initial conditions of the control variables for the trajectories displayed on \Cref{fig:Pareto}, as well as the final state control variables.  %Ideally, the final state should be the same regardless of the initial conditions. 
%In terms of rotational speed, the agent consistently aims to achieve approximately 11 rpm, with discrepancies likely attributable to the discrete nature of the actions. %There is a higher variance over the final pitch value, this likely occurs because the training might not have fully converged.  
%No puedes decir que no ha convergido.. porque te van a decir pues convergelo

Overall, the agent robustness has been tested on a simple steady wind scenario. The agent is able to reduce the wind turbine noise to admissible levels while maximizing power. Furthermore, the agent finds optimum operational conditions  regardless of the initial condition, showing the robustness of the algorithm. In other words, the neural network which approximates $Q(s,a)$ has covered all his input space $\mathcal S \times \mathcal A$. 

\subsection{Control Strategy for Experimental Winds}
\label{subsec:ExpControl}
The agent capabilities are now tested over experimental wind conditions. We compare the energy extraction between our agent and two controllers that are designed solely to maximize power. By doing so, we can demonstrate how much power we need to sacrifice to keep the wind turbine at acceptable decibel levels. The performance of the three controllers are going to be compared. These controllers include:
\begin{itemize}
    \item \textbf{Classic wind turbine controller}: Standard wind turbine controller designed to reach the power curve of the wind turbine, using torque or pitch control depending on whether the wind speed is above or below rated wind speed. Details can be seen in \ref{appendix:windturbinecontroller}.
    \item \textbf{\textit{Power} DDQN}: Agent designed to maximize solely power. It is trained with no noise penalization, that is only the power reward is included, see  \cref{eq:power_reward}. %with the boundaries barrier function reward are used.
    \item \textbf{\textit{Quiet} DDQN}: Agent designed to maximize power without producing more that 45 dB decibel levels at 100 m downwind of the rotor. It is trained with the complete reward definitions including power and noise, see \cref{eq:complete_reward}. 
\end{itemize}

\begin{figure}[htbp]
    \begin{subfigure}[b]{0.4\textwidth}
        \centering
        \includegraphics[width=6.5cm]{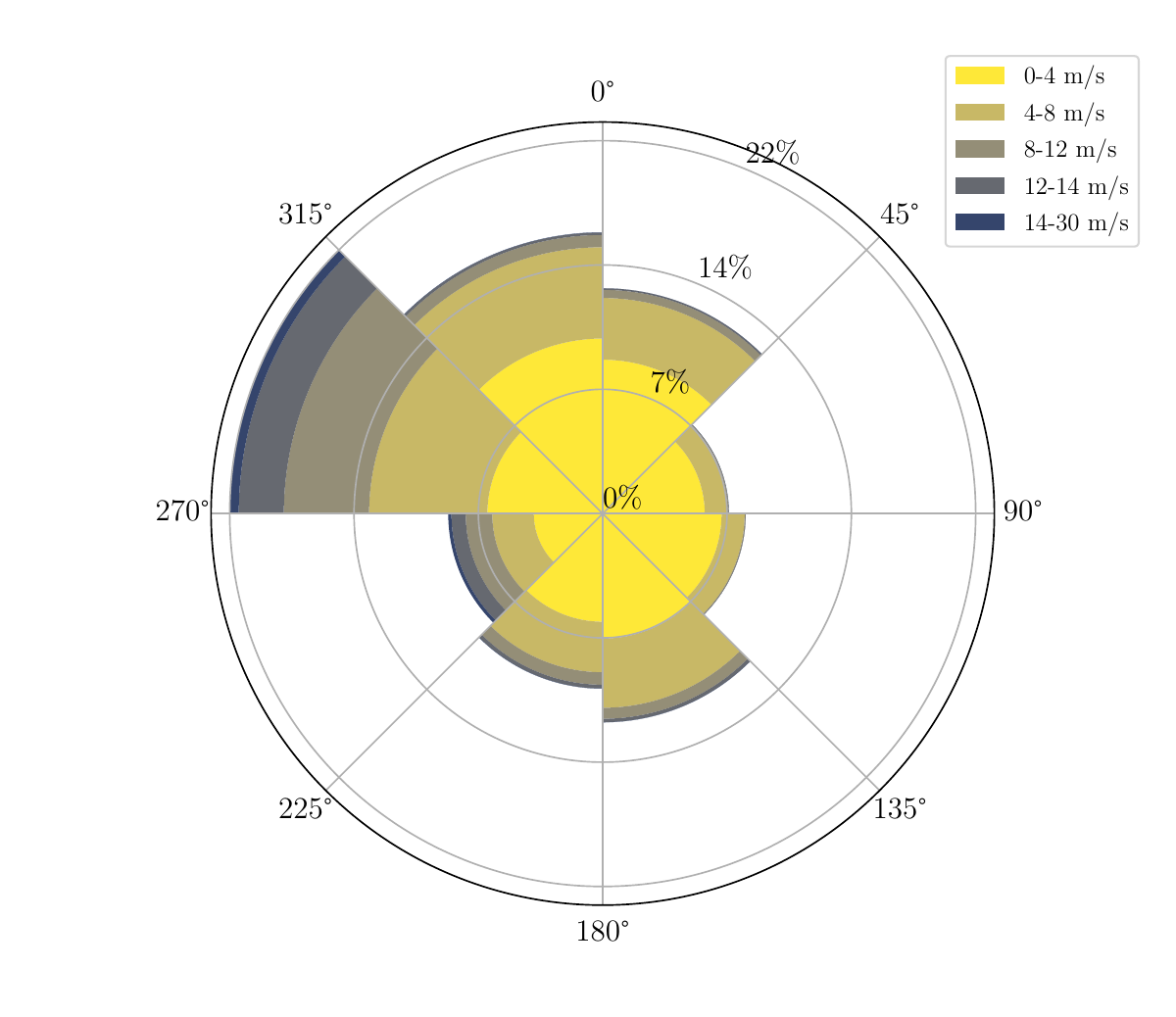}
        \caption{Wind rose for one year of data.}
        \label{fig:windrose}
    \end{subfigure}
    \hspace{1.5cm}
    \begin{subfigure}[b]{0.4\textwidth}
        \centering
        \includegraphics[width=8cm]{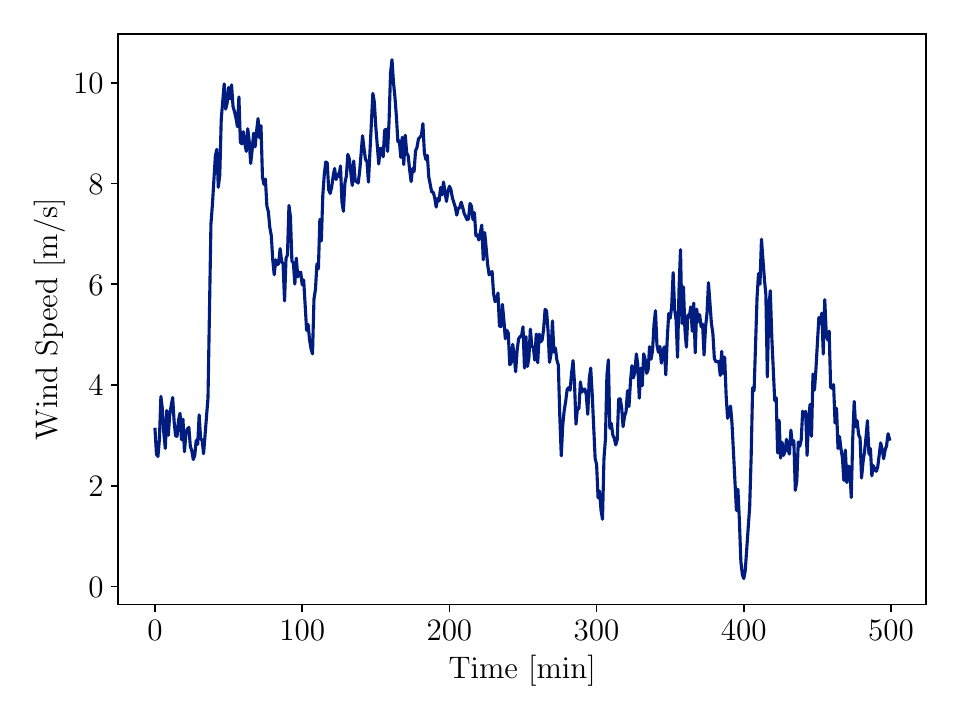}
        \caption{Wind speed distribution for 8 hours.}
        \label{fig:exp_ws}
    \end{subfigure}
    \caption{Wind conditions on the \textit{experimental wind environment}.}
    \label{fig:exp_wind}
\end{figure}

The wind data used to validate the control performance under real wind conditions was obtained from the measurement and instrumentation data center (MIDC) of NREL, see \cite{JAA96}. These daily wind measurements are available as open-source. For this study, wind speed and wind direction measurements at an 80 m height from June 1, 2023, to June 1, 2024, are selected. \Cref{fig:windrose} displays a wind rose illustrating the wind speed and wind direction of this dataset. Since this study concentrates on torque-pitch control, it is assumed that the incoming wind is consistently aligned with the wind turbine, a condition typically managed by yaw control. Therefore, we assume perfect alignment and only the wind speed distribution is used in the subsequent results.       

Note that the \textit{Power} DDQN agent considers power optimization uniquely. Therefore, it is only applicable to the below rated wind speed region defined on \ref{appendix:windturbinecontroller}. When the wind speed exceeds the rated value, the control strategy maintains nominal power rather than maximizing it. To achieve this behavior with a reinforcement learning agent, the reward function would need to be modified. Consequently, the wind speed distribution used to validate the agent under turbulent wind conditions is restricted to below-rated wind speeds, enabling a meaningful comparison between the three controllers. 

The control performance of the three agents is analyzed in detail over an 8-hour time span, using the wind speed distribution shown in  \Cref{fig:exp_ws}. The RL controllers are allowed to control each minute. In the next section, we will estimate the annual energy production for all controllers. 

\begin{figure}[htbp]
    \centering
    \begin{subfigure}[b]{0.45\textwidth}
        \centering
        \includegraphics[width=\textwidth]{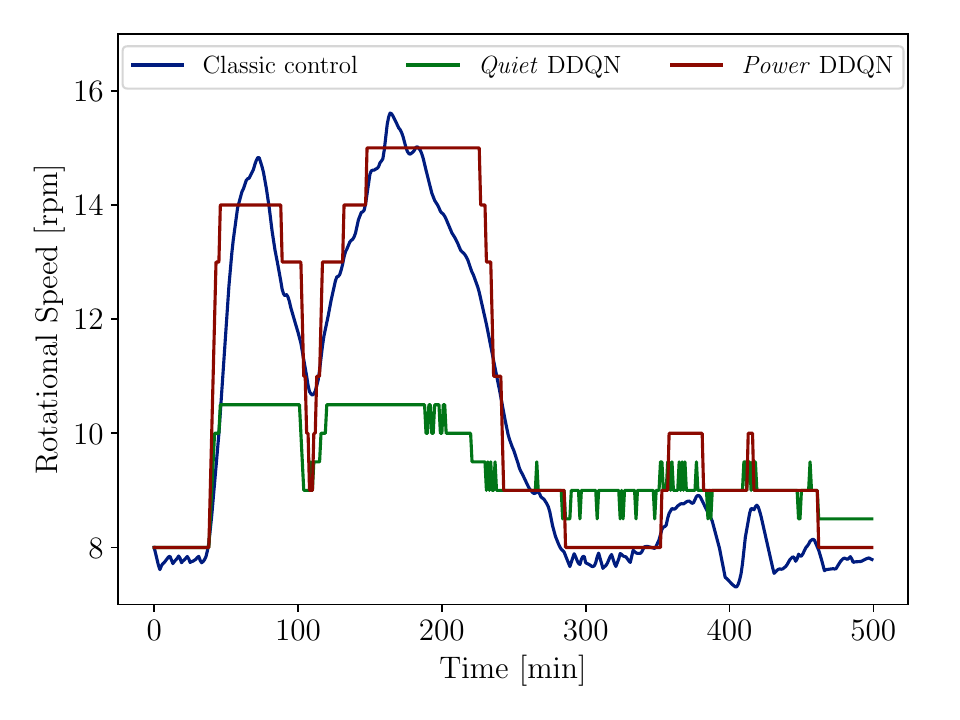}
        \caption{Rotational speed control.}
        \label{fig:exp_rpm}
    \end{subfigure}
    \hfill
    \begin{subfigure}[b]{0.45\textwidth}
        \centering
        \includegraphics[width=\textwidth]{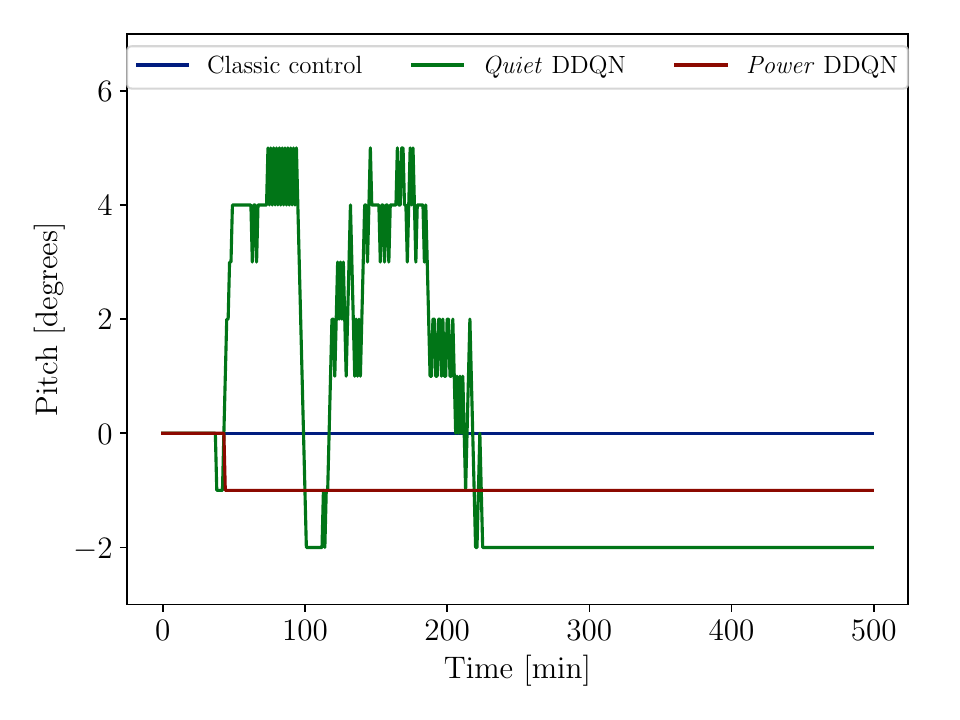}
        \caption{Pitch control.}
        \label{fig:exp_pitch}
    \end{subfigure}
    \\
    \begin{subfigure}[b]{0.45\textwidth}
        \centering
        \includegraphics[width=\textwidth]{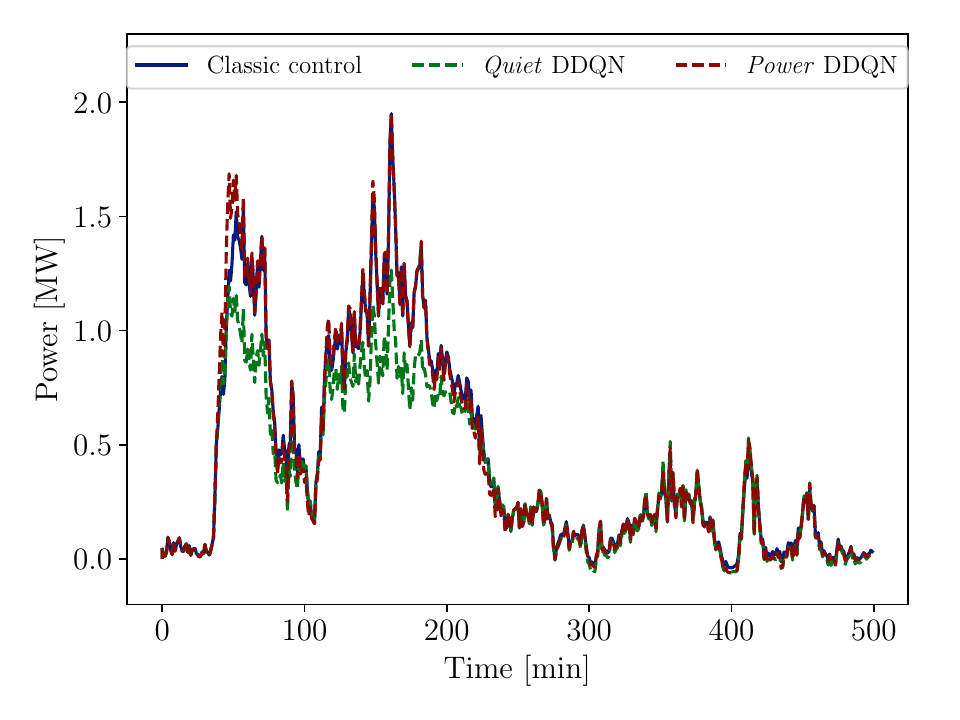}
        \caption{Power generation.}
        \label{fig:exp_power}
    \end{subfigure}
    \hfill
    \begin{subfigure}[b]{0.45\textwidth}
        \centering
        \includegraphics[width=\textwidth]{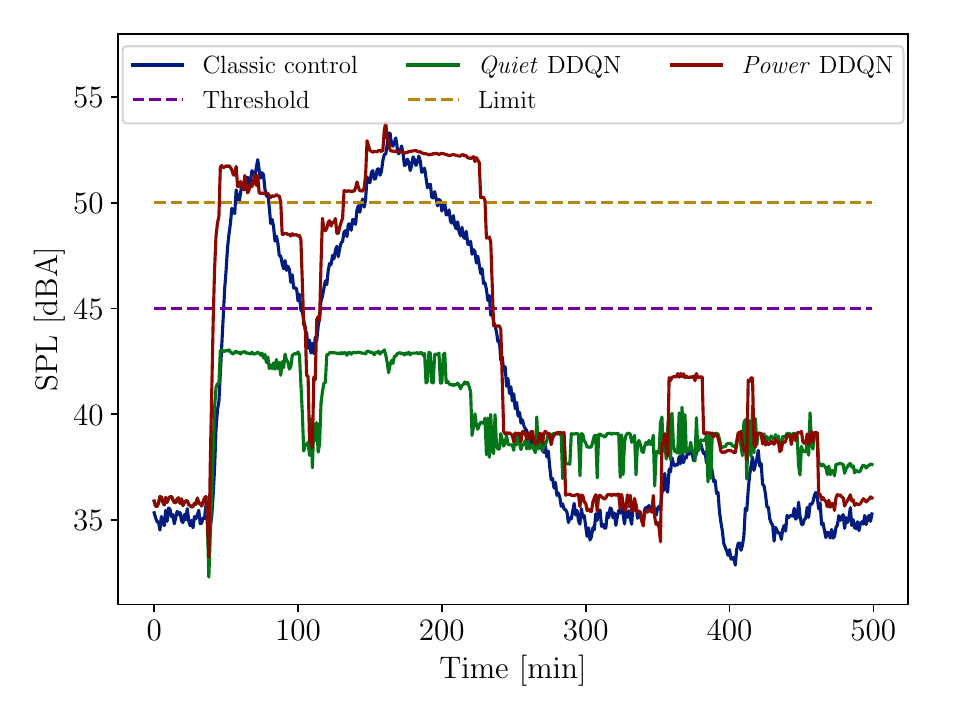}
        \caption{Sound Pressure Level of the wind turbine.}
        \label{fig:exp_spl}
    \end{subfigure}
    \caption{Control results for the three different agents on the 8 hour \textit{experimental wind environment}.}
    \label{fig:exp_results}
\end{figure}

\Cref{fig:exp_results} shows the results from the different controllers over the first 8 hours of the yearly dataset. \Cref{fig:exp_rpm,fig:exp_pitch} display the control parameters evolution while \Cref{fig:exp_power,fig:exp_spl} illustrates the power and the noise generated 100 m downwind. It is noted that the \textit{Power} agent matches the power extraction achieved by the classic control strategy, essentially implementing the same control approach but with the discrete actions defined for the reinforcement learning agent. Since neither of these controllers is designed to consider the acoustics of the wind turbine, both generate high levels of noise when the wind speed is sufficiently high. In contrast, the \textit{Quiet} agent can match the power generation of the power-oriented controllers when the wind speed is moderate. When wind speeds are higher, it extracts as much power as possible while keeping noise levels below the threshold value. Moreover, all three controllers maintain a constant pitch value to maximize power extraction.  However, the \textit{Quiet} agent adjust the pitch angle to reduce the noise levels when the wind speeds get higher. 

This test shows the flexibility of the RL strategy for control and highlights the possibility of including multi-objectives. In addition, we see that there is no need to have an \textit{a priori} knowledge of the turbine performance (e.g., the power curve or rated maximum power) since the RL will learn these characteristics when trained.

\subsection{Annual wind energy estimation}
\begin{figure}[htbp]
    \centering
    \includegraphics[width=7cm]{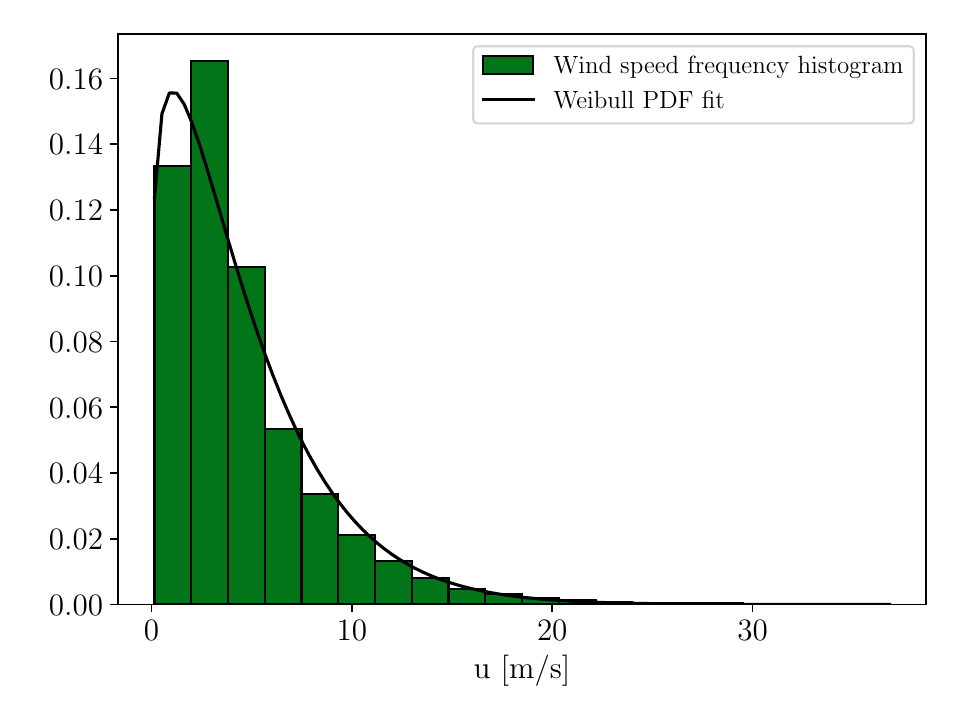}
    \caption{Wind speed frequency histograms of one year experimental data and PDF of the adjusted Weibull distribution. The shape and scale parameters are $k = 1.195$ and $c=4.837$ respectively.}
    \label{fig:weibull}
\end{figure}

There are different methodologies to obtain an estimate of the annual generation of wind energy, \citep{garcia2009comparison}. The standard procedure is based on decoupling the wind turbine from the wind distribution of the particular site. It considers the observed wind speed frequency histogram to fit a theoretical probability density function (PDF) for the wind speed. It also requires a transfer function that models the relation between power output and wind speed. 
Typically, the Weibull distribution is used to fit the wind speed frequency histogram. \cite{garcia2008influence} showed that although the Weibull distribution may not be substantiated for most sites, it does not include important errors on the energy estimations. The probability density function of the Weibull distribution is given by:
\begin{equation}
    \label{eq:Weibull}
    f_U(u) = \left(\frac kc\right)\cdot\left(\frac uc\right)^{k-1}\exp{\left(-\left(\frac uc\right)^k\right)} ,
\end{equation}
where $U$ denotes the random variable that models the wind speed. The fit of the shape and scale parameters $k$ and $c$ are established from the mean and variance of the wind speed, $\mu_U=\mathbb E[U]$ and $\sigma_U^2 = \mathbb V[U]$. The specific relations can be seen in \cite{spiru2024wind} work and are the following: 
\begin{equation}
    k = \left(\frac{\sigma_U}{\mu_U}\right)^{-1.086},
\end{equation}
and 
\begin{equation}
    c = \frac{\mu_U}{\Gamma\left(1+\frac 1k\right)},
\end{equation}
where $\Gamma$ denotes the special gamma function. 
This formulation can be employed to obtain the Weibull probability density function that represents the one-year experimental data reported by \cite{JAA96}. This is illustrated in \Cref{fig:weibull}. 

\begin{figure}[htbp]
    \centering
    \begin{subfigure}[b]{0.45\textwidth}
        \centering
        \includegraphics[width=\textwidth]{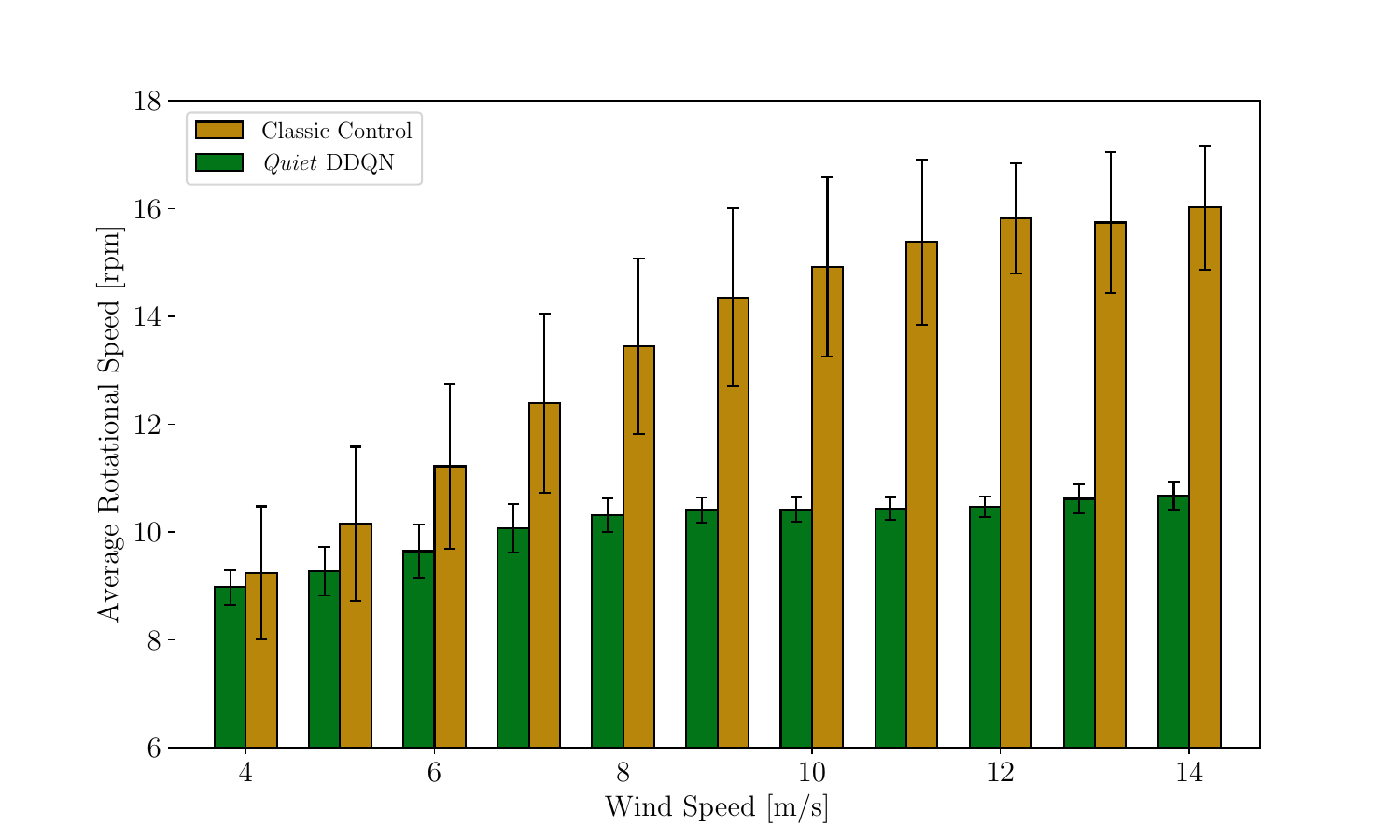}
        \caption{Rotational speed empiric operational curve.}
        \label{fig:EPC_rpm}
    \end{subfigure}
    \hfill
    \begin{subfigure}[b]{0.45\textwidth}
        \centering
        \includegraphics[width=\textwidth]{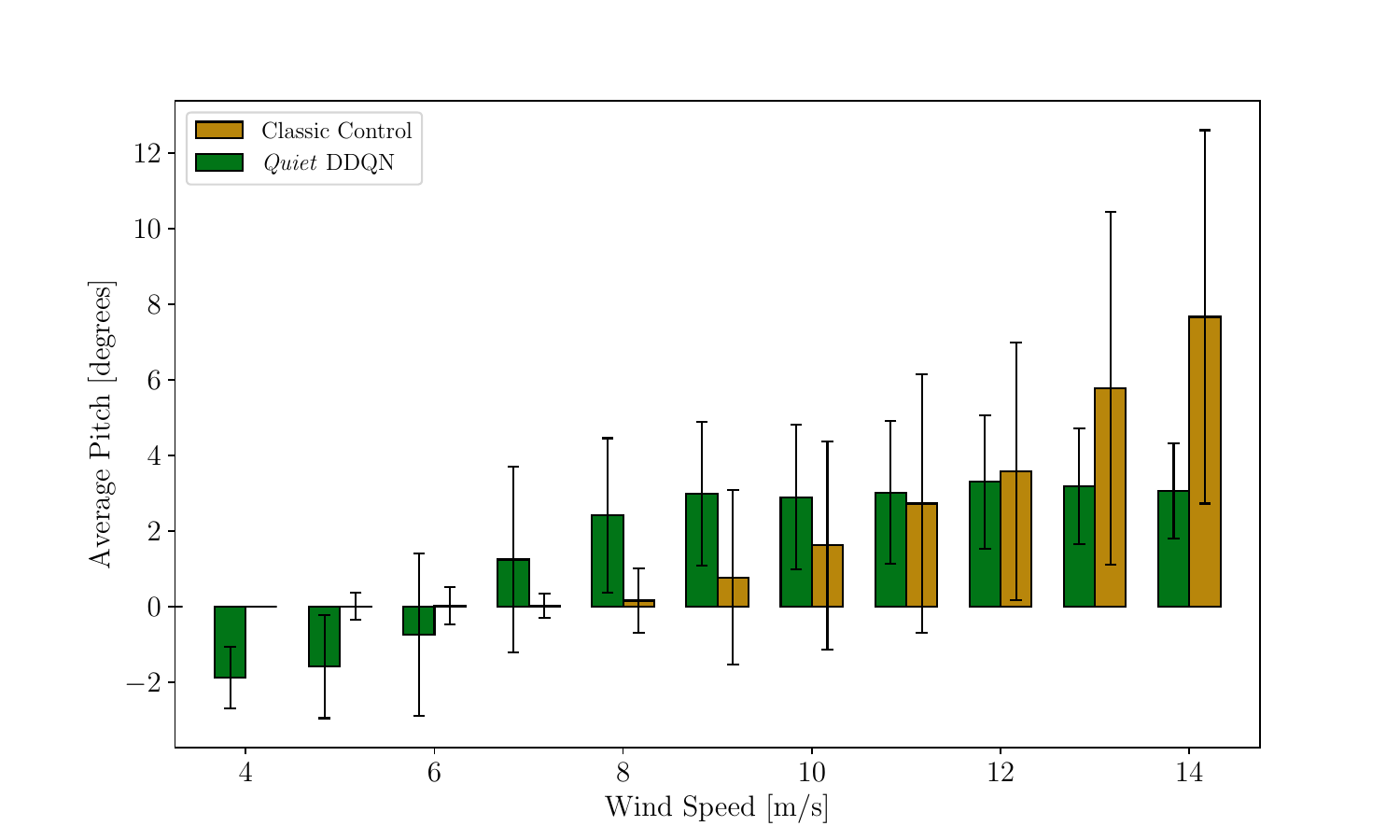}
        \caption{Pitch empiric operational curve.}
        \label{fig:EPC_pitch}
    \end{subfigure}
    \\
    \begin{subfigure}[b]{0.45\textwidth}
        \centering
        \includegraphics[width=\textwidth]{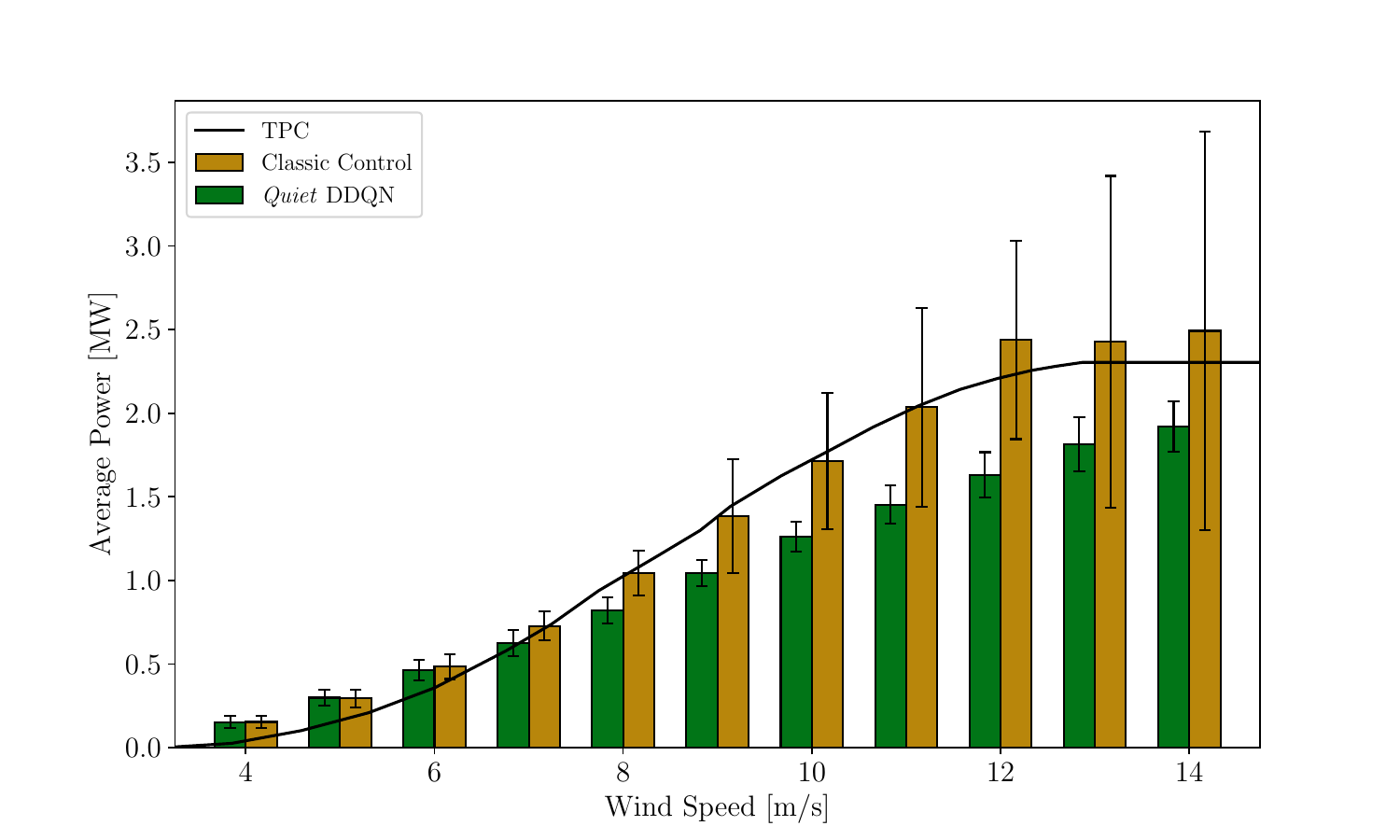}
        \caption{Effective Power Curve (EPC) of both control strategies against the Theoretical Power Curve (TPC) from the manufacturer.}
        \label{fig:EPC_power}
    \end{subfigure}
    \hfill
    \begin{subfigure}[b]{0.45\textwidth}
        \centering
        \includegraphics[width=\textwidth]{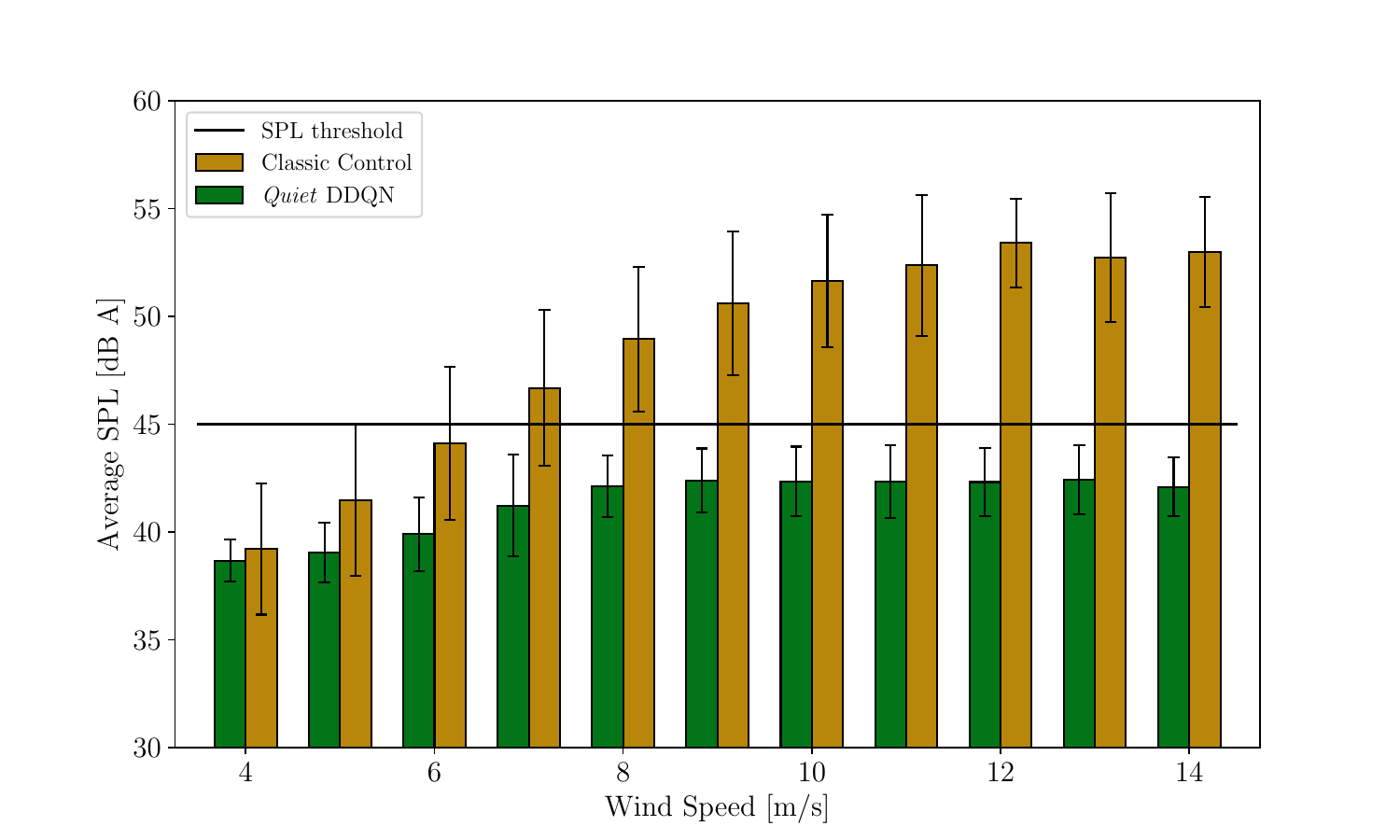}
        \caption{Sound Pressure Level of the wind turbine observed at 100 m downwind for both control strategies against the SPL threshold.}
        \label{fig:EPC_spl}
    \end{subfigure}
    \caption{Average values of the control variables for wind speed bins, obtained from simulating two different control agents over a 100 hour experimental wind speed period. The error bars denote the standard deviation.}
    \label{fig:EPC_results}
\end{figure}

Regarding the transfer function between power and wind speed, various strategies exist \citep{abolude2018assessment}. The Theoretical Power Curve (TPC) does not account for control mechanisms. Furthermore, since our wind turbine control strategy considers acoustic generation, the wind turbine will exhibit a significantly different Effective Power Curve (EPC). It is necessary to compute an EPC that accurately represents the transfer function between power and wind speed for our specific control scenario.
The EPC can be computed using simulations of the wind turbine control. The agent is faced against a turbulent wind that covers all the range of interest of wind speed, mainly between cut-in and cut-off wind speed. This turbulent wind must be representative of the turbulent nature of the wind that the wind turbine is going to face during operation. Once the simulation is done, all the pairs of data points $(U_\infty,C_p)$ can be used to obtain a transfer function for the power coefficient $C_p(U_\infty)$. 
A subset of 100 hours of the experimental wind measurements from the MIDC \citep{JAA96} has been used to obtain the EPC of the SWT2.3-93 wind turbine using the Classic Control and the \textit{Quiet} DDQN agent already introduced on \Cref{subsec:ExpControl}. \Cref{fig:EPC_results} illustrates the results of this simulations, showing the operational laws of control for each agent on \Cref{fig:EPC_rpm,fig:EPC_pitch} and the SPL and power associated on \Cref{fig:EPC_spl,fig:EPC_power} respectively. The behavior is as expected, the \textit{Quiet} Agent does not increase the rotational speed above 10.5 rpms to avoid surpassing the SPL threshold and uses the pitch to reduce noise if needed, which explains the high variance bars on the pitch (see \cref{fig:EPC_pitch}) and low ones in the rotational speed (see \cref{fig:EPC_rpm}). Meanwhile the classic control can increase the rotational speed freely and the pitch is only use in above-rated wind speed scenarios, see \ref{appendix:windturbinecontroller}. The large standard deviations on the classic control pitch are due to the PID control, which is dynamically adjusting to the turbulent wind. In \Cref{fig:EPC_power} it is illustrated how the classic control matches on average the TPC. However, it is not able to adjust perfectly to the turbulent wind, showed by its high variance on the above-rated wind speed region. The \textit{Quiet} agent achieves less power than the classic one but is able to maintain the sound pressure level below the specified threshold of 45 dB A, see \cref{fig:EPC_spl}.

\begin{figure}[htbp]
    \centering
    \begin{subfigure}[b]{0.45\textwidth}
        \centering
        \includegraphics[width=\textwidth]{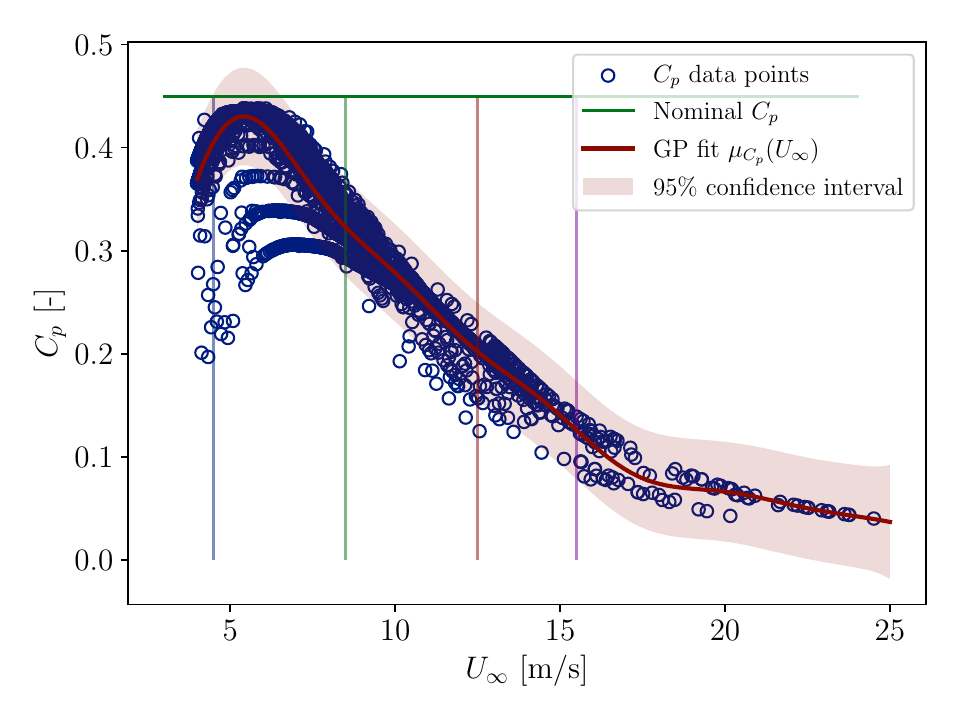}
        \caption{Power coefficient EPC. The data points represent all the observed $(U_\infty,C_p)$ pairs observed during the simulation performed to obtain the EPC. The Gaussian Process Regression fit, mean and confidence interval is included.}
        \label{fig:EPC_Cp}
    \end{subfigure}
    \hfill
    \begin{subfigure}[b]{0.45\textwidth}
        \centering
        \includegraphics[width=\textwidth]{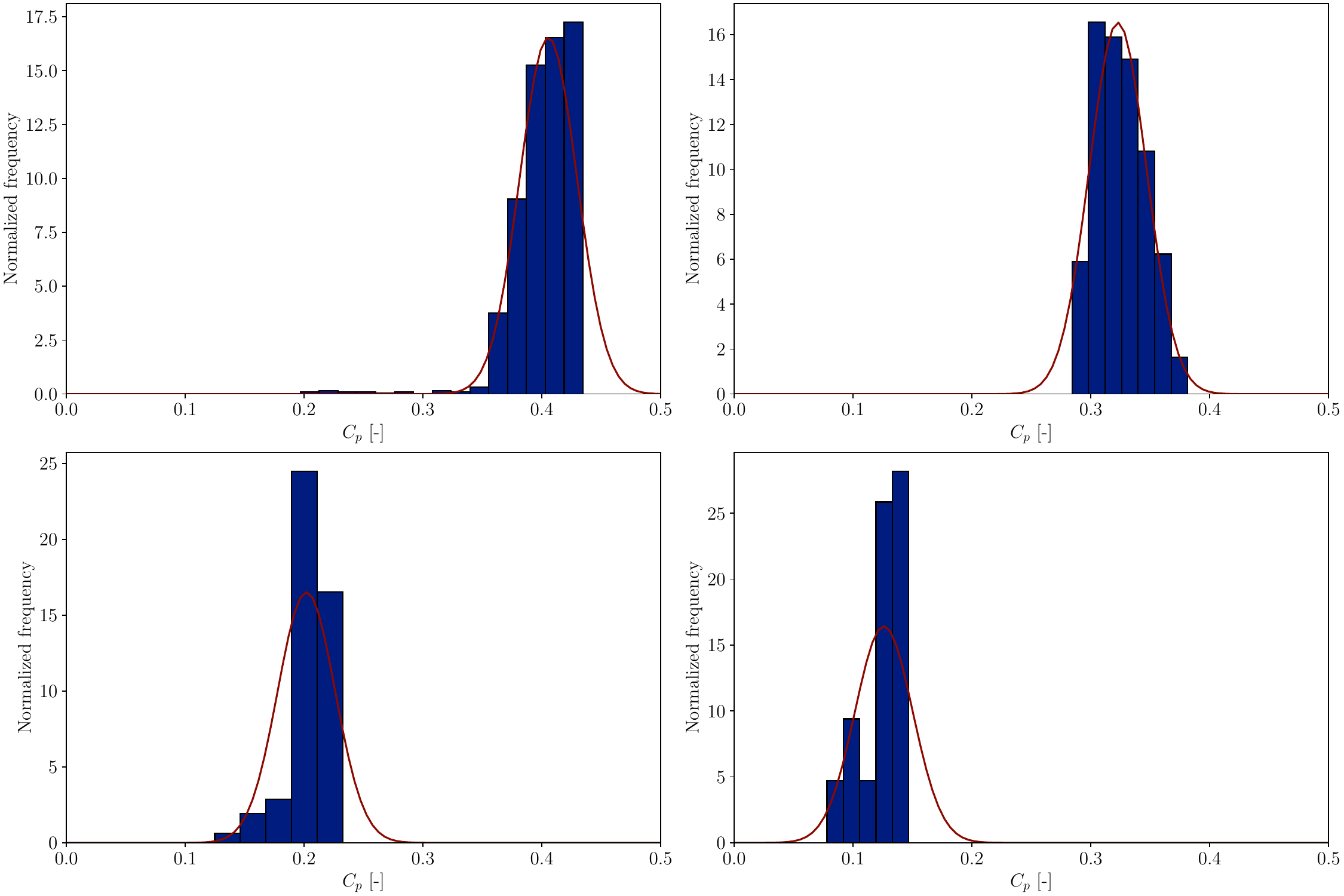}
        \caption{Comparison of the $C_p$ frequency histograms and the Gaussian probability distribution (\textcolor{python_red}{\raisebox{0.5ex}{\rule{1em}{1pt}}}) obtained with the GP for different wind speed values. The wind speed values for each histogram are specified in \Cref{fig:EPC_Cp} with vertical colored lines. The order is the following:\\ Top Left: \transparency[0.5]{\textcolor{python_blue}{\rule{1.25ex}{1.75ex}}} , Top Right: \transparency[0.5]{\textcolor{python_green}{\rule{1.25ex}{1.75ex}}} , Bottom Left: \transparency[0.5]{\textcolor{python_red}{\rule{1.25ex}{1.75ex}}}, Bottom Right: \transparency[0.5]{\textcolor{python_purple}{\rule{1.25ex}{1.75ex}}}.}
        \label{fig:Cp_distributions}
    \end{subfigure}
\caption{Power coefficient obtained from simulating the \textit{Quiet} DDQN controller over a 100 hour experimental wind speed period.}
\label{fig:CP_DISTRIBUTION}
\end{figure}

Traditional approaches use only the average value or polynomial fits of the historical/simulated data to construct the EPC. All these methods do not capture the variance of the data in the model. To account for this variability on the EPC model we introduce a statistical method. For simplicity, we model the power coefficient $C_p(U_\infty)$, which is obtained by non-dimensionalizing the EPC data. A Gaussian Process Regression \citep{mackay1998introduction} can be employed to model the power coefficient at each wind speed as a Gaussian probability distribution, $C_p(U_\infty)\sim\mathcal N[\mu_{C_p}(U_\infty),\sigma_{C_p}(U_\infty)]$. \Cref{fig:EPC_Cp} shows the power coefficients points obtained after the wind turbine control simulation. This data is used to fit the Gaussian Process (GP) model and obtain the mean and standard deviation of the power coefficient as functions of the wind speed, this fit is also included in \Cref{fig:EPC_Cp}. The $C_p$ distribution for specific values of the wind speed is illustrated on \Cref{fig:Cp_distributions} where it is compared with the histogram of the power coefficient from the data.

Assuming the Weibull probability distribution for the wind speed $U_\infty$ combined with the GP model for the $C_p$ distribution, the estimation of the annual wind energy can be performed by computing the expectation of the wind energy $E_w$. Mathematical details on the statistical distributions are provided in \ref{appendix:statistics}.
\Cref{tab:controllers} presents the annual energy estimation for each control strategy. It is important to note that the \textit{Power} DDQN controller is applicable only in the below-rated wind speed region. Therefore, when computing the annual wind energy generation with this control, we impose nominal power for wind speeds above the rated value. The \textit{Power} DDQN controller is included in the comparison to ensure that it remains competitive with the Classical Control within the below-rated wind speed range. The \textit{Quiet} DDQN controller is able to control the wind turbine without surpassing the sound pressure level threshold selected and obtains an 78\% of the annual energy production compared to the Classical Control. Additionally, in \Cref{tab:controllers} it is shown the average standard deviation across the wind speed for the power coefficient GP fit. There, it is shown that the \textit{Quiet} agent exhibits the control strategy with the least variance. 

\begin{table}[htbp]
    \centering
    \begin{tabular}{c|c|c}
        \hline
        \textbf{Controller} & \(\mathbf{E_w}\) \textbf{[MWh]} & \(\mathbf{\hat\sigma_{C_p}}\) \\
        \hline
        \textit{Quiet} DDQN & 2722 & 0.024 \\
        \textit{Power} DDQN & 3541 & 0.042 \\
        Classic Controller & 3458 & 0.057 \\
        \hline
    \end{tabular}
    \caption{Annual wind energy generation for the three different controllers. Additionally, the average standard deviation across wind speed for each controller is presented, $\hat\sigma_{C_p} = \int_{U_{\text{in}}}^{U_\text{off}}\sigma_{C_p}(u)f_U(u)du$, details on \ref{appendix:statistics} at \cref{eq:variance_expectation}.} 
    \label{tab:controllers}
\end{table}

\section{Conclusions}
\label{sec:conclusions}
In conclusion, integration of reinforcement learning with wind turbine control holds promise for optimizing energy generation and efficiency while minimizing acoustic environmental impact.
A DDQN reinforcement learning agent can replicate the control strategy of a standard wind turbine controller without prior explicit knowledge of the wind turbine, relying solely on a wind turbine solver for experiential learning. Moreover, advanced control strategies can be readily implemented by modifying the reward function. In this work, an RL controller is defined to maximize power output while maintaining acceptable decibel levels, thereby incorporating acoustic effects into the control strategy. This demonstrates that MORL is capable of dynamically balancing two different objectives effectively.

An effective power curve is computed from control simulations of turbulent wind data. This allow to characterize the reinforcement learning control strategy, obtaining the operational laws and obtaining an annual wind energy estimation. The methodology is validated using a SWT2.3-93 wind turbine with
a rated power of 2.3 MW. We evaluate the yearly energy production for a realistic site. The DDQN reinforcement learning control provides similar energy production that a traditional control. The methodology presented allows for the inclusion of noise limits leading to a 22\% reduction in the annual energy extraction when activating a maximum allowed noise of 45 dB (100 meters downwind downwind of the turbine).

Further research directions include investigating Multi-Agent Reinforcement Learning algorithms for cooperative control of wind turbines within farms, which could enhance overall system performance while controlling noise at the farm level.    
\newpage\appendix 
\section{Statistical details for the EPC model}
\label{appendix:statistics}
Let us consider a two dimension random variable $\mathbb X = (C_p,U)$. This random variable model the probability of obtaining a certain wind speed with a certain power coefficient. The wind speed marginal distribution accounts for the global wind speed distribution of the localization of the wind turbine. Meanwhile, the power coefficient distribution measures the performance of the wind turbine at different wind speeds. 

There exist an \textit{a priori} unknown joint probability density $f(c_p,u)$. The power generated by the wind turbine, $P$, is a function of this random variable, so it is itself a random variable, 
$$ P = \frac12\rho A C_pU^3. $$

The wind energy, $E_w$, that the wind turbine extracts from the wind for a given period can be written as $E_w = \int_0^T P dt$. However, this statistical model does not include information about the temporal evolution of $C_p$ and $U$. We can compute the expectation of the wind energy using the expectation of the power over a period of time.
\begin{equation}
\label{eq:raw_expectation}
    \mathbb E[E_w] = \Delta T\mathbb E[P] = \Delta T \int \frac12\rho A c_pu^3 f(c_p,u) dc_p du,
\end{equation}
where $\Delta T$ denotes the period of time. Notice that this only make sense if the  unknown wind speed evolution $U_\infty(t)$ fits in the annual distribution modeled by the random variable $U$.  

Although we do not know the joint PDF, we know that the wind speed random variable $U$ follows a Weibull distribution. Therefore, the marginal probability density function of $U$, $f_U(u)$ is a weibull PDF that follows \cref{eq:Weibull}. 

On the other hand we can obtain the distribution of power coefficient for each wind speed value. This would be the conditioned power coefficient PDF, that is $f_{C_p}(c_p|U=u)$. From this two PDF we can obtain the joint PDF, using the following relation:

\begin{equation}
\label{eq:conditioned_prob}
    f_{C_p}(c_p\mid U=u) = \frac{f(c_p,u)}{f_U(u)}
\end{equation}
In this work, the conditional power coefficient PDF is obtained using a Gaussian Process Regression algorithm. Therefore, its density function is the following: 
\begin{equation}
    f_{C_p}(c_p\mid U=u) = \frac1{\sqrt{2\pi\sigma_{C_p}(u)^2}}\exp{\left(\displaystyle\frac{c_p-\mu_{C_p}(u)}{\sigma_{C_p}(u)}\right)^2},
\end{equation} 
where the mean $\mu_{C_p}(u)$ and standard deviation $\sigma_{C_p}(u)$ are obtained from the wind turbine control simulations. 

Finally, the expectation of the wind energy can be compute as follows: 
\begin{equation}
    \mathbb E[E_w] = \Delta T\mathbb E[P] = \left(\frac12\Delta T\rho A\right) \int_{U_{\text{in}}}^{U_{\text{off}}}\left(\int_0^{C_{p,\text{nom}}} c_pf_{C_p}(c_p|U=u)dc_p\right) u^3 f_U(u) du.
\end{equation}
Notice that the inner integral is the expectation of the conditional distribution, $\mathbb E[C_p|U=u] = \mu_{C_p}(u)$. Hence, the expectation of the wind energy only requires the mean of the distribution fitted by the GP. The variance of the control, $\sigma_{C_p}(u)^2$, has no influence on the estimation of the wind energy. However, it gives us information about the control and can be useful to measure, this can be done computing the expectation of the variance. 
\begin{equation}
\label{eq:variance_expectation}
    \mathbb E[\sigma_{C_p}(U)] = \int_{U_{\text{in}}}^{U_\text{off}}\sigma_{C_p}(u)f_U(u)du
\end{equation}
\newpage

\section{Standard Wind Turbine Control Strategy}\label{appendix:windturbinecontroller}
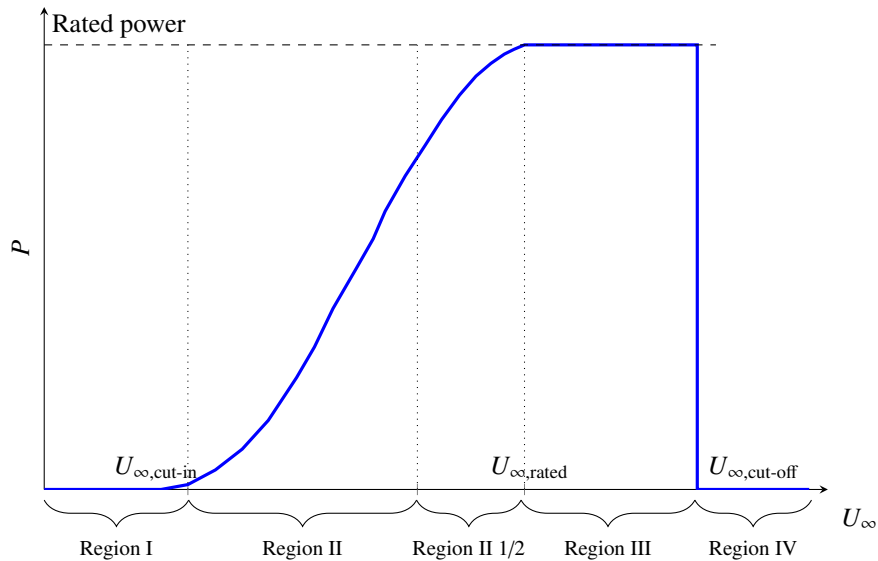
\begin{figure}[htbp]
    \centering
    \begin{tikzpicture}
    \begin{axis}[
        width=12cm,
        height=8cm,
        xlabel={$U_\infty$},
        ylabel={$P$},
        xlabel style={anchor=east, xshift=6cm},
        xmin=0, xmax=21,
        ymin=0, ymax=2500,
        xtick=\empty,
        ytick=\empty,
        extra x ticks={3.85, 10.00, 12.87},
        extra x tick labels= \empty,
        grid=none,
        axis lines=left,
    ]
    
    % Datos de la curva de potencia
    \addplot[very thick,blue] coordinates {
        (0.0, 0.0)
        (3.132, 0.0)
        (3.85, 25.35)
        (4.588, 101.4)
        (5.306, 209.1)
        (6.004, 358.0)
        (6.762, 579.8)
        (7.241, 738.3)
        (7.739, 937.9)
        (8.358, 1144.0)
        (8.816, 1299.0)
        (9.136, 1442.0)
        (9.674, 1625.0)
        (10.21, 1784.0)
        (10.65, 1917.0)
        (11.13, 2044.0)
        (11.57, 2142.0)
        (11.97, 2208.0)
        (12.33, 2256.0)
        (12.59, 2281.0)
        (12.87, 2304.0)
        (13.36, 2304.0)
        (13.84, 2304.0)
        (14.3, 2304.0)
        (14.76, 2304.0)
        (17.00, 2304.0)
        (17.50, 2304.0)
        (17.50, 0.0)
        (20.50, 0.0)
    };
    
    % Línea de potencia nominal
    \draw[dashed] (axis cs:0,2304) -- (axis cs:18,2304);
    \node at (axis cs:2,2400) {Rated power};
    
    % Líneas de lambda
    \draw[dotted] (axis cs:3.85,0) -- (axis cs:3.85,2304);
    \draw[dotted] (axis cs:10.0,0) -- (axis cs:10.0,2304);
    \draw[dotted] (axis cs:12.87,0) -- (axis cs:12.87,2304);

    \node[above] at (13,0) {$U_{\infty,\text{rated}}$};
    \node[above] at (3,0) {$U_{\infty,\text{cut-in}}$};
    \node[above] at (19,0) {$U_{\infty,\text{cut-off}}$};
    \end{axis}

    % Agregar corchetes y etiquetas de regiones
    \draw [decorate,decoration={brace,amplitude=10pt,mirror,raise=4pt}] (0,0) -- (1.92,0) node [black,midway,yshift=-0.8cm] {\footnotesize Region I};
    \draw [decorate,decoration={brace,amplitude=10pt,mirror,raise=4pt}] (1.92,0) -- (4.95,0) node [black,midway,yshift=-0.8cm] {\footnotesize Region II};
    \draw [decorate,decoration={brace,amplitude=10pt,mirror,raise=4pt},] (4.95,0) -- (6.34,0) node [black,midway,yshift=-0.8cm] {\footnotesize Region II 1/2};
    \draw [decorate,decoration={brace,amplitude=10pt,mirror,raise=4pt}] (6.34,0) -- (8.65,0) node [black,midway,yshift=-0.8cm] {\footnotesize Region III};
    \draw [decorate,decoration={brace,amplitude=10pt,mirror,raise=4pt}] (8.65,0) -- (10.2,0) node [black,midway,yshift=-0.8cm] {\footnotesize Region IV};

\end{tikzpicture}
    \caption{Control strategy regions for a variable-speed, horizontal-axis wind turbine on the power curve.}
    \label{fig:pid_control_strategy}
\end{figure}

The control strategy for variable-speed horizontal-axis wind turbines can be divided into four regions based on wind speed. Although each region definition may vary depending on the specific control design, the fundamental objectives within each region are as follows:
\begin{itemize}
    \item \textbf{Region I}: When the wind speed is below the cut-in value, the turbine cannot operate.
    \item \textbf{Region II}: At wind speeds above the cut-in threshold but below the rated speed, the primary objective is to optimize power generation. This is achieved by adjusting the rotor speed to align with the power curve of the wind turbine, utilizing a predetermined lookup table.
    \item \textbf{Region III}: When wind speeds exceed the rated value, the focus shifts to maintaining a consistent rotor speed across a broad range of wind velocities. This is typically achieved through adjustment of the blade pitch, commonly implemented using a proportional-integral-derivative (PID) control strategy, although there are more sophisticated approaches \citep{8511531}. 
    \item \textbf{Region IV}: When the wind speed surpasses the cut-off value, the turbine must be shut down for safety.
\end{itemize}
The transition between regions II and III, sometimes referred to as Region II $\frac12$, is characterized by maintaining a constant rotor speed. Although there are different options depending on the specific control design. \Cref{fig:pid_control_strategy} illustrates these regions on the power curve of the wind turbine. Further details on classical wind turbine control strategies can be found in the works of \cite{burton2011wind} or \cite{85c4d1c00a4b43a7a2aad5174c969473}. 

The controller module in OpenFAST facilitates the customization of controllers. In this study, the wind turbine controller is derived from the OpenFAST implementation from \cite{mulders2018delft}, tailored to suit the characteristics of the SWT wind turbine. 

\section*{Acknowledgments}
Esteban Ferrer and Oscar A. Marino would like to thank the support of
Agencia Estatal de Investigación for the grant "Europa Excelencia" for the project EUR2022-134041 funded by MCIN/AEI/10.13039/501100011033) and the European Union NextGenerationEU/PRTR and also the funding received by the Grant DeepCFD (Project No. PID2022-137899OB-I00) funded by MICIU/AEI/10.13039/501100011033 and by ERDF, EU.  
This research has been cofunded by the European Union (ERC, Off-coustics, project number 101086075). Views and opinions expressed are, however, those of the author(s) only and do not necessarily reflect those of the European Union or the European Research Council. Neither the European Union nor the granting authority can be held responsible for them.
%Finally, all authors gratefully acknowledge the Universidad Politécnica de Madrid (www.upm.es) for providing computing resources on the Magerit Supercomputer.
\bibliographystyle{plainnat}
\bibliography{biblio}
%\printbibliography[heading=bibintoc] % Print the bibliography
\end{document}